\documentclass[a4paper,12pt]{article}

\usepackage[margin=1in,letterpaper]{geometry} 
\usepackage{graphicx} 
\usepackage[final]{hyperref}
\usepackage{authblk}
\hypersetup{
	colorlinks=true,
	linkcolor=blue,
	citecolor=red,
	filecolor=magenta,
	urlcolor=blue         
}
\usepackage{amsmath,amssymb,amsthm}
\usepackage{caption}
\usepackage{subcaption}
\usepackage[capitalize,nameinlink,noabbrev,nosort]{cleveref}
\usepackage{tikz}
\usepackage{pgfplots}
\usetikzlibrary{fadings}
\pgfplotsset{compat=1.18}

\newcommand{\ds}{\displaystyle}
\newcommand{\qbinom}[2]{\bgroup\renewcommand*{\arraystretch}{1}\begin{bmatrix} #1 \\ #2\end{bmatrix} \egroup}
\newcommand{\qbinomt}[2]{\bgroup\renewcommand*{\arraystretch}{1}\begin{bmatrix} #1 \\ #2\end{bmatrix}_{\!t} \egroup}

\let\deg\relax
\DeclareMathOperator{\deg}{deg}
\DeclareMathOperator{\ord}{ord}

\newcommand{\wt}{\text{wt}_{\rf,t}}
\newcommand{\wta}{\text{wt}_{a,t}}
\newcommand{\avg}[1]{\langle #1 \rangle}
\newcommand{\asip}{$(q, t, \rf)$~ASIP}
\newcommand{\bt}[1]{[#1]_t}
\newcommand{\rf}{\theta} 
\newcommand{\pa}{a} 
\newcommand{\bbt}[3]{\varphi_{#1,#3}(#2)} 
\newcommand{\dt}{\bullet}
\newcommand{\s}{\mid}
\newcommand{\Za}{Z^a}

\begin{document}
\pagenumbering{arabic}
\title{An exactly solvable asymmetric simple inclusion process}
\author[$\dagger$]{Arvind Ayyer}
\author[$\star$]{Samarth Misra}
\affil[$\dagger$]{Department of Mathematics, 
Indian Institute of Science, Bangalore  560012, India.}
\affil[$\star$]{Department of Physics, 
Indian Institute of Science, Bangalore  560012, India.}
\date{\today}
\maketitle

\begin{abstract}
We study a generalization of the asymmetric simple inclusion process (ASIP) on a periodic one-dimensional lattice, where the integers in the particles rates are deformed to their $t$-analogues. We call this the \asip{}, where $q$ is the asymmetric hopping parameter and $\rf$ is the diffusion parameter.
We show that this process is a misanthrope process, and consequently the steady state is independent of $q$.
We compute the steady state, the one-point correlation and the current in the steady state.
In particular, we show that the single-site occupation probabilities follow a 
\emph{beta-binomial} distribution at $t=1$.
We compute the two-dimensional phase diagram in various regimes of the parameters $(t, \rf)$ and perform simulations to justify the results.
We also show that a modified form of the steady state weights at $t \neq 1$ satisfy curious palindromic and antipalindromic symmetries.
Lastly, we define an enriched process at $t=1$ and $\rf$ an integer which projects onto the $(q, 1, \rf)$~ASIP and whose steady state is uniform, which may be of independent interest.
\end{abstract}

\section{Introduction}
The asymmetric simple exclusion process (ASEP) in one-dimensions is an extremely well-studied interacting particle system both in statistical physics and in mathematics. 
It is important from the point of view of nonequilibrium statistical physics because it is exactly solvable and explicit calculations have led to a lot of insight into nonequilibrium phenomena in one dimension.
It has also turned out to be of great interest in different areas of mathematics such as combinatorics, probability theory and representation theory.
In the ASEP, every site has at most one particle and particles hop preferentially onto neighbouring sites provided they are empty. Therefore, one can think of it as a `fermionic' process. And indeed, like in fermionic statistics, particles in the ASEP do tend to repel each other.
{The ASEP is an example of an integrable model, and there are generalizations  such as the $q$-TASEP with modified rates that are also integrable~\cite{Alexei2014}.
}

It is natural to consider a `bosonic' counterpart of the ASEP, and a symmetric analog known as the \emph{symmetric inclusion process} (SIP)
was first introduced and studied in its own right by Giardin\`a--Redig--Vafayi~\cite{giardina-redig-vafayi-2010} from the point of view of obtaining correlation inequalities. 
We add that a model very similar in spirit is implicit in the 
works of Giardin\`a--Kurchan--Redig~\cite[Section~III]{giardina-kurchan-redig-2007} and 
Giardin\`a--Kurchan--Redig--Vafayi~\cite[Section~5.2]{giardina-kurchan-redig-vafayi-2009}, where they obtain it as the dual of a system with Brownian interactions.
Unlike the ASEP, the inclusion process permits multiple particles per site
and the dynamics promotes aggregation. Roughly speaking, if two neighbouring sites have $a$ and $b$ particles, the rate in the SIP at which a particle moves from the first to the second is $a( \rf + b)$ and from the second to the first is $b( \rf + a)$, where $\rf$ is a free parameter, called the \emph{diffusion parameter} in the literature.

The asymmetric inclusion process (ASIP), first proposed by Grosskinsky--Redig--Vafayi \cite{grosskinsky-etal-2011}, is a natural variant of the SIP, where particles hop preferentially in one direction. They studied the ASIP in the one-dimensional lattice with closed boundaries, and it was extended to two variants of the ASIP with periodic boundary conditions by Cao--Chleboun--Grosskinsky~\cite{cao-etal-2014}.
Both these works study condensation phenomena both in and out of equilibrium in certain limits of the rates. A lot of work has been done since then on condensation in the ASIP. We refer to the survey by Landim~\cite{landim-2019} for more details. To be more precise, \cite{cao-etal-2014} studied two variants of the ASIP. The first was precisely the SIP (with symmetric hopping rates), and the second was
with totally asymmetric hopping rates, which they call the TASIP.
We simultaneously generalize both the models in this work. Specifically, we generalize the form of the rates mentioned above to $[a]_t (\rf + [b]_t)$, where $[n]_t$ is the $t$-analogue of the integer $n$, and we add an asymmetry parameter $q$; see \cref{sec:model} for the precise definition. We call this model the \emph{\asip{}}. In the limit $t \to 1$, we obtain both the variants studied in \cite{cao-etal-2014,Jatuviriyapornchai2020} at $q = 0$ and $q = 1$. 
To clarify, we only focus on the steady state.
We will show that many of the properties of the ASIP continue to hold for the \asip{}.
{We however, note that unlike the $q$-TASEP mentioned above, we do not know
if the \asip{} is integrable. We have verified that it is not obviously integrable in the sense that the local Markov matrices do not satisfy the Yang--Baxter equation.}

After the definition of the \asip{} in \cref{sec:model}, we will show that this is a special case of a \emph{misanthrope process}~\cite{cocozza-1985,evans-waclaw-2014}, and so the steady state will be of product form. 
In \cref{sec:ss}, we give explicit formulas for the steady states and 
derive some properties.
It will turn out that the analysis for $t = 1$ and $t \neq 1$ 
will be different, and these will be studied separately throughout.
We look at observables in the steady state in \cref{sec:obs}. In particular, we
will show that the one-point distribution is the so-called beta-binomial distribution (which generalizes the beta distribution) and calculate the current when $t = 1$.

In \cref{sec:phase diagram}, we derive the phase diagram of the \asip{} in terms of the parameters $\rf$ and $t$. Although we are unable to give exact results, we explain the phases in various limiting regions of the diagram. 
{In particular, we find a region of the phase diagram which suggests
\emph{hyperuniformity}~\cite{jack-etal-2015}, where the variance in the density profile is highly suppressed. Such behavious has been recently found in other processes of the type we consider here~\cite{chleboun-et-al-2018}.}
We also attach movies as ancillary files together with this submission in these regions and include snapshots of these movies. In some cases, the simulation results do not seem to match calculations, and we explain why in each of these cases.

{When $t \neq 1$, we obtain a new symmetry of the stationary distribution.
First, we use an alternate parameterization of the rates by replacing 
$\rf$ by a different parameter $a$, and show that the steady state weights are either palindromic or antipalindromic polynomials jointly in the variables $a, t$ in \cref{sec:palindromicity and antipalindromicity} with the same center of mass. 
As a result, we obtain that the stationary distribution is invariant
under the transformation $a \to 1/a, t \to 1/t$.
This is similar in spirit to a result we had obtained earlier for the $(q, t)$~$K$-ASEP~\cite{ayyer-misra-2024} under the transformation $t \to 1/ t$}.
Lastly, in \cref{sec:enriched}, we construct an enriched process when $t = 1$ and $\rf$ is a positive integer which projects as a Markov process onto the \asip{}. We show that the steady state of this enriched process is uniform, and thus obtain an alternate proof of the steady state formula when $t = 1$.

\section{Model description}
\label{sec:model}

We define an asymmetric simple inclusion process, denoted \asip{}, on a periodic one-dimensional lattice characterized by the following parameters: the number of sites $L \in \mathbb{N}$ (sites $1$ and $L$ are adjacent) and the total number of particles $n \in \mathbb{N}$, the asymmetry parameter $q \geq 0$ which distinguishes the forward and backward transition rates, and diffusion parameter $\rf > 0$ appearing in the target site contribution (to which the particle hops), and the deformation parameter $t \geq 0$. All particles are indistinguishable and can occupy any site. 
We denote the set of all configurations by
\begin{equation}
\Omega_{L,n} = \left\{ \eta  = (\eta_1,\eta_2,\ldots,\eta_L) \in \{0,1,\dots,n\}^{L} \Biggm|  \sum_{i=1}^L \eta_i = n \right\}.
\end{equation}
The total number of configurations $\bigm| \Omega_{L,n} \bigm|$ is thus given by the number of ways of distributing $n$ particles among $L$ sites, or the number of compositions of $n$ into $L$ non-negative parts giving
\begin{equation}
    \bigm| \Omega_{L,n} \bigm| = \binom{n+L-1}{n}.
\end{equation}
To illustrate this, we can take a small example with $L=3$ sites and $n=4$ particles, giving us 
\begin{equation}
\label{example small all configs}
\Omega_{3,4} = 
\begin{Bmatrix}
(0,0,4), (0,1,3), (0,2,2), (0,3,1), (0,4,0),\\
(1,0,3), (1,1,2), (1,2,1), (1,3,0), (2,0,2),\\
(2,1,1), (2,2,0), (3,0,1), (3,1,0), (4,0,0) 
\end{Bmatrix}
\end{equation}
with $| \Omega_{3,4}| =\binom{6}{4} = 15$.

\begin{figure}[h!]
\centering
\begin{subfigure}{.5\textwidth}
  \centering
  \includegraphics[width=.7\linewidth]{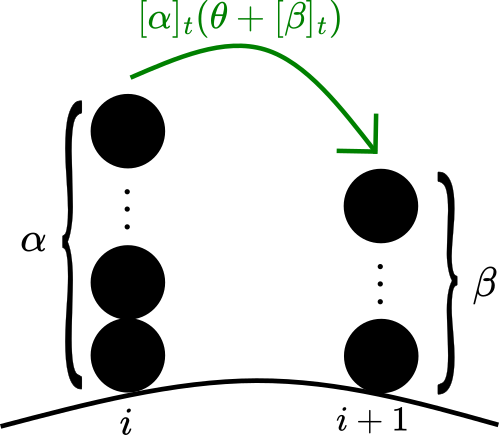}
  \caption{Forward transition rate}
  \label{fig:sub1}
\end{subfigure}%
\begin{subfigure}{.5\textwidth}
  \centering
  \includegraphics[width=.7\linewidth]{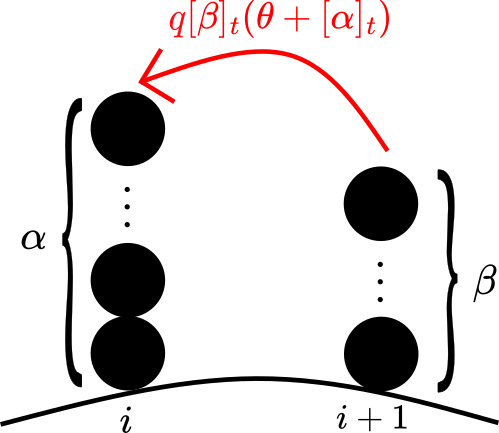}
  \caption{Backward transition rate}
  \label{fig:sub2}
\end{subfigure}
\caption{Transition rates between two consecutive sites indexed by $i$ and $i+1$ in clockwise order having particles $\alpha$ and $\beta$ respectively.}
\label{fig:rates}
\end{figure}

To define the rates of the \asip{}, we recall the \emph{$t$-analogue of a nonnegative integer $k$} as
\begin{equation}
[k]_t = \frac{1-t^k}{1-t} = 
\begin{cases}
\ds \sum_{i = 0}^{k - 1} t^i & \text{for } t \neq 1,\\
k & \text{for } t=1.
\end{cases}
\end{equation}
For later purposes, we define the \emph{$t$-factorial} of a nonnegative integer as
\begin{equation}
[k]_t! = [k]_t [k-1]_t \cdots [1]_t,
\end{equation}
and the \emph{$t$-binomial coefficient} or \emph{Gaussian polynomial} as 
\begin{equation}
\qbinomt{m}{n} = \frac{[m]_t!}{[n]_t! [m-n]_t!},
\end{equation}
where $m$ and $n$ are nonnegative integers with $n \leq m$.
The \asip{} is a simple process, meaning that particles can only hop between neighbouring sites. {Only one particle will make a jump in a single transition step.} We will denote configurations by
$\eta = (\eta_1,\eta_2,\ldots,\eta_L) \in \Omega_{L,n}$,
where $\eta_i$ denotes the number of particles at site $i$, also known as the \emph{occupation number}. 
The transitions are as follows.
For two neighbouring sites indexed $(i,i+1)$ having occupation numbers $(\alpha,\beta)$ respectively, 
\begin{equation}
\label{forward rates t}
(\alpha, \beta) \rightarrow (\alpha-1, \beta +1 ) \quad \text{with rate} \quad \bt{\alpha} (\rf + \bt{\beta})
\end{equation}
for the forward transition, and
\begin{equation}
\label{backward rates t}
(\alpha, \beta) \rightarrow (\alpha + 1, \beta -1 ) \quad \text{with rate} \quad q \, \bt{\beta} (\rf + \bt{\alpha})
\end{equation}
for the reverse transition.
A special case of our model, namely $t = 1$, coincides with a special case of the model studied by Grosskinsky--Redig--Vafayi~\cite[Section 3]{grosskinsky-etal-2011}.
These rates are automatically $0$ when the source site is empty, so a transition out of an empty site is forbidden. See \cref{fig:rates} for an illustration.

To establish the relationship between the \asip{} and the broader class of misanthrope processes, we verify the fundamental constraints. Recall that a \textit{misanthrope process} is a simple process in which the transition rate for a neighbouring pair of sites containing $(\alpha, \beta)$ particles to transition to $(\alpha-1, \beta+1)$ is $u(\alpha,\beta)$. It has been shown~\cite{cocozza-1985,evans-waclaw-2014} that when the rates satisfy the 
conditions
\begin{equation}
\label{misanthrope1}
\frac{u(\beta, \alpha)}{u(\alpha+1, \beta-1) } = \frac{u(1, \alpha) u(\beta, 0)}
{u(\alpha+1, 0) u(1, \beta-1)},
\end{equation}
and
\begin{equation}
\label{misanthrop2}
u(\beta, \alpha) - u(\alpha, \beta) = u(\beta, 0) - u(\alpha, 0)
\end{equation}
we get a product form for the steady state which is independent of $q$.
For the totally asymmetric ASIP (i.e. the $(0, t, \rf)$~ASIP), we have the rate
\begin{equation}
    u(\alpha,\beta) = \bt{\alpha} (\rf + \bt{\beta}),
\end{equation}
and the left-hand side of \eqref{misanthrope1} is
\begin{equation}
    \frac{u(\beta, \alpha)}{u(\alpha+1, \beta-1) } = \frac{\bt{\beta}(\rf+\bt{\alpha})}{\bt{\alpha+1}(\rf+\bt{\beta-1})},
\end{equation}
which is equal to the right-hand side
\begin{equation}
    \frac{u(1, \alpha) u(\beta, 0)}{u(\alpha+1, 0) u(1, \beta-1)} = \frac{(\rf+\bt{\alpha})\bt{\beta}\rf}{\bt{\alpha+1}\rf (\rf+\bt{\beta-1})}.
\end{equation}
Similarly, the left-hand side of \eqref{misanthrop2} is
\begin{equation}
     u(\beta,\alpha)-u(\alpha,\beta) = \bt{\beta}(\rf+\bt{\alpha}) - \bt{\alpha}(\rf+\bt{\beta}) = \rf (\bt{\beta}-\bt{\alpha}),
\end{equation}
which is equal to the right-hand side
\begin{equation}
    u(\beta,0) - u(\alpha,0) = \bt{\beta} \rf - \bt{\alpha} \rf = \rf (\bt{\beta}-\bt{\alpha}).
\end{equation}
This verification confirms that the $(0, t, \rf)$~ASIP belongs to the class of misanthrope processes, and thus the steady state is of product form. 
{We note in passing that there are generalizations of the misanthrope process
studied in the literature to which our process also belongs~\cite{Chleboun2013}.}
By standard arguments, it is also clear that the steady state of the \asip{} is the same as that of the $(0, t, \rf)$~ASIP and hence, independent of $q$.

When $t \neq 1$, we also parametrize our rates differently in terms of
\begin{equation}
    \rf = \frac{1-\pa}{\pa(1-t)},
\end{equation} 
so that the transitions depend on $\pa$ and $t$. In this notation, 
\begin{equation}
\label{forward rate a and t}
(\alpha, \beta) \rightarrow (\alpha-1, \beta +1 ) \quad \text{with rate} \quad \bt{\alpha} 
\left( \frac{1-\pa}{\pa(1-t)} + \bt{\beta} \right) = \frac{\bt{\alpha}(1-\pa t^\beta)}{\pa(1-t)},
\end{equation}
and 
\begin{equation}
\label{backward rate a and t}
(\alpha, \beta) \rightarrow (\alpha + 1, \beta -1 ) \quad \text{with rate} \quad q \, \bt{\beta} 
\left( \frac{1-\pa}{\pa(1-t)} + \bt{\alpha} \right) = \frac{q\bt{\beta}(1-\pa t^\alpha)}{\pa(1-t)}.
\end{equation}
We will focus on this formulation only while discussing palindromic symmetry in \cref{sec:phase diagram}.

\section{Steady state}
\label{sec:ss}

Having established that the \asip{} belongs to the class of misanthrope processes, we now derive the exact form of the steady state. We begin with the general values of $t$ before specializing to $t = 1$.

\subsection{$t$ general}
For $q \geq 0$ and $\rf > 0$, it is easy to show that there is a sequence of transitions leading from any configuration in $\Omega_{L,n}$ to any other. This proves that the process is ergodic. 
Note that if $\rf = 0$, no particle can enter an empty site, and ergodicity is broken. Hence, we require $\rf > 0$. 
We focus on properties of the steady state, which is unique by ergodicity and which we denote by $\pi$. 
Thus, the probability of seeing any configuration $\eta = (\eta_1,\eta_2,\ldots,\eta_L) \in \Omega_{L,n}$ in the long-time limit approaches $\pi(\eta)$. 
Using the result for the product state of a misanthrope process outlined in \cite[Equation (22)]{evans-waclaw-2014},  
\begin{equation}\label{eq: evans eqn 1}
    f(m) = \prod_{i = 1}^{m} \frac{u(1, i-1)}{u(i, 0)},
\end{equation}
and $\pi(\eta) \propto \prod_i f(\eta_i)$. For the \asip{}, this quantity is
\begin{equation}\label{eq: evans eqn 2}
     f(m) = \prod_{i = 1}^{m} \frac{\rf + \bt{i-1}}{\rf \bt{i}}.
\end{equation}

Define the function $\bbt{\rf}{m}{t}$ by
\begin{equation}\label{eq:defining box bar t}
    \bbt{\rf}{m}{t} = \prod_{i=0}^{m-1} (\rf + \bt{i}),
\end{equation}
with $\bbt{\rf}{0}{t}=1$.  We can rewrite \eqref{eq: evans eqn 2} as
\begin{equation}
\label{density}
f(m) = \frac{\bbt{\rf}{m}{t}}{\rf^{m}\bt{m}!},
\end{equation}
We thus get the steady state weights using~\cite[Equation (4)]{evans-waclaw-2014} to be
\begin{equation}
    \wt(\eta) = \prod\limits_{i=1}^L f(\eta_i)
    = \prod\limits_{i=1}^L \frac{\bbt{\rf}{\eta_i}{t}}{\rf^{\eta_i}\bt{\eta_i}!}.
\end{equation}
By dropping the constant scaling factor of $\prod\limits_{i=1}^{L} \rf^{\eta_i}= \rf^{n}$ from the denominator and multiplying the numerator by a constant $\bt{n}!$, we write the weights as
\begin{equation}\label{wt_eta1}
\wt(\eta) = \qbinomt {n}{\eta_1,\ldots,\eta_L} \prod_{i=1}^L \bbt{\rf}{\eta_i}{t}
\end{equation}
where
\begin{equation}
    \qbinomt {n}{\eta_1,\ldots,\eta_L} = \frac{\bt{n}!}{\bt{\eta_1}!\bt{\eta_2}!\ldots\bt{\eta_L}!}
\end{equation}
is the usual $t$-multinomial coefficient. 
As mentioned before, the steady state is of product form. 
Hence, weights of configurations depend only on the content 
and are independent of the ordering of sites.
We can write its steady state probability as
\begin{equation}
    \pi(\eta) = \frac{\wt(\eta)}{Z_{L,n}} = \frac{1}{Z_{L,n}} \qbinomt {n}{\eta_1,\ldots,\eta_L} \prod_{i=1}^L \bbt{\rf}{\eta_i}{t}
\end{equation}
where
\begin{equation}
\label{pf sum}
Z_{L,n} = \sum_{\eta \in \Omega_{L,n}} \wt(\eta)
\end{equation}
is the \emph{nonequilibrium partition function} which normalizes the probability distribution. 
The reader can verify that detailed balance holds at $q=1$, but not otherwise.
The weight function defined in \eqref{wt_eta1} is a bivariate polynomial of $\rf,t \in \mathbb{R^{+}} \setminus \{0\}$. As an example, the weights for the process with $L=3$ and $n=4$ are
\begin{equation}
\label{eg:L=3,n=4,t factorized}
\begin{aligned}
\wt(0,0,4) &= \rf (\rf + 1) ( \rf + 1 + t)( \rf + 1 + t + t^2), \\
\wt(0,1,3) &= (1 + t) (1 + t^2) \, \rf^2(\rf+1)( \rf + 1+t),\\
\wt(0,2,2) &= (1 + t^2) (1 + t + t^2)\, \rf^2 (\rf+1)^2\\
\wt(1,1,2) &= (1 + t) (1 + t^2) (1 + t + t^2)\, \rf^3(\rf + 1).  
\end{aligned}
\end{equation}

{It is known that for $t=1$ and $\rf=1$, the steady-state distribution is uniform \cite{grosskinsky-etal-2011}. We generalize this result and prove that it remains valid under $\rf=1/t$. When} $\rf = 1/t$, we obtain 
\begin{equation}
\bbt{1/t}{m}{t} 
  = \frac{1}{t} \left( \frac{1}{t}+\bt{1} \right) \cdots 
  \left(\frac{1}{t}+\bt{m-1} \right)
= \frac{\bt{m}!}{t^{m}}
\end{equation}
so the weight function \eqref{wt_eta1} becomes
\begin{equation}
    \textrm{wt}_{1/t, t}(\eta) = \frac{\bt{n}!}{t^n}
\end{equation}
which is the same for every configuration. 
Thus, the uniform distribution is the steady state in this case,
\begin{equation}\label{eq: uniform when theta t = 1}
    \pi(\eta) = \frac{1}{\bigm| \Omega_{L,n} \bigm|} = \frac{1}{\ds\binom{L+n-1}{n}}.
\end{equation}

\subsection{$t=1$}
\label{sec: t=1}
We now turn our attention to the special case when $t=1$, where the model simplifies considerably. 
{As mentioned in the introduction, this case at $q = 0$ and $q = 1$ has  been
studied in \cite{cao-etal-2014,Jatuviriyapornchai2020} 
where they use the notation $d_L$ for what we call $\rf$. 
Another case with rightward drift $0<q<1$ has been studied in \cite{grosskinsky-etal-2011}}.
The transition rates discussed previously in \eqref{forward rates t} reduce to 
\begin{equation}
\label{forward rate t=1}
(\alpha, \beta) \rightarrow (\alpha-1, \beta +1 ) \quad \textrm{with rate} \quad \alpha\,(\rf + \beta),
\end{equation}
and the reverse transition \eqref{backward rates t} becomes
\begin{equation}
\label{backward rate t=1}
(\alpha, \beta) \rightarrow (\alpha+1, \beta-1 ) \quad \textrm{with rate} \quad q\,\beta\,(\rf + \alpha).
\end{equation}
The $t$-multinomial coefficient also simplifies to the regular multinomial coefficient
\[
\bgroup
\begin{bmatrix} n 
\\ \eta_1,\eta_2,\ldots,\eta_L
\end{bmatrix}_{\!t=1}
\egroup = \binom{n}{\eta_1,\eta_2,\ldots,\eta_L}.
\]
To express the weights in a more familiar form, we recall the \emph{rising factorial} 
\begin{equation}\label{rising_factorial}
    x^{\overline{r}} =\frac{\Gamma(x+r)}{\Gamma(x)} = \frac{(x+r-1)!}{(x-1)!} = x \,(x+1) \dots (x+r-1),
\end{equation}
for any real number $x$. 
We can thus simplify the steady state weight \eqref{wt_eta1} to
\begin{equation}    
\label{weight s.s. t=1}
   \textrm{wt}_{\rf,1} = \binom{n}{\eta_1,\dots, \eta_L} \prod_{i=1}^{L} \rf^{\overline{\eta_i}}
\end{equation}
and the partition function is
\begin{equation}
Z_{L,n} = \sum_{\eta \in \Omega_{L,n}} \textrm{wt}_{\rf,1}
\end{equation}
For example, when $L=3$ and $n=4$, substitute $t=1$ in \eqref{eg:L=3,n=4,t factorized} to obtain
\begin{equation}\label{example Z t=1}
    Z_{3,4} = \sum_{\eta \in \Omega_{3,4}} 
    \textrm{wt}_{\rf,1}(\eta) =81\rf^4 + 162\rf^3 + 99\rf^2 + 18\rf.
\end{equation}

We now give a remarkable closed-form expression for the partition function
when $t=1$.
Recall the rising factorial variant of the \emph{Chu--Vandermonde identity}~\cite[Equation {[13d]}]{comtet-1974}, which states that
\begin{equation}
    \sum^{n}_{i=0} \binom{n}{i} x^{\overline{i}} y^{\overline{n-i}} = (x+y)^{\overline{n}}.
\end{equation}
Similarly, we have the analogous multinomial variant,
\begin{equation}
\label{multi-chu-van}
\sum_{\eta_1+\eta_2+\ldots+\eta_L=n} \binom{n}{\eta_1, \eta_2, \ldots, \eta_L} x_1^{\overline{\eta_1}} x_2^{\overline{\eta_2}} \ldots x_L^{\overline{\eta_L}} = (x_1 + \ldots + x_L)^{\overline{n}}.
\end{equation}
Comparing this with our partition function for the weight function \eqref{weight s.s. t=1}, we obtain
\begin{equation}
\label{pf t=1}
\begin{split}
Z_{L,n} =& \sum_{\eta_1+\cdots+\eta_L=n} \binom{n}{\eta_1,\dots, \eta_L}  \prod_{i=1}^{L} \rf^{\overline{\eta_i}} \\
=& (L \rf)^{\overline{n}}
= L \rf (L \rf + 1) \ldots (L \rf + n -1) .
\end{split}
\end{equation}
For our previous example, we obtain
\[
Z_{3,4} = (3\rf)^{\overline{4}} = 3\rf(3\rf+1)(3\rf+2)(3\rf+3) = 81 \rf^4 + 162 \rf^3 +99 \rf^2 +18 \rf,
\]
which matches the formula in \eqref{example Z t=1}.

\section{Observables}
\label{sec:obs}

We now calculate observables in the steady state of this process.

\subsection{One-point correlation}
Since multiple particles can occupy a single site, we can compute the distribution of the number of particles at a given site. We denote the steady state probability that there are $\alpha$ particles at site $i$ by $\avg{\eta_i = \alpha}$.

Since the steady state weights in \eqref{wt_eta1} are of product form, we can easily show that
\begin{equation}
\label{s.s. prob eta_i = alpha}
    \avg{\eta_i= \alpha} = \qbinomt{n}{\alpha} \bbt{\rf}{\alpha}{t}  \frac{Z_{L-1,n-\alpha}}{Z_{L,n}},
\end{equation}
for any site $i$. This is not easy to compute for general $\rf$ and $t$. But for $t=1$, the formulas become much simpler. Using \eqref{pf t=1}, we get
\begin{equation}
\label{s.s. prob eta_i = alpha t=1}
    \avg{\eta_i= \alpha} =  \binom{n}{\alpha}
    \rf^{\overline{\alpha}} \frac{((L-1)\rf)^{\overline{n-\alpha}}}{(L \rf)^{\overline{n}}}.
\end{equation}
This formula is also given in \cite[Page~526]{cao-etal-2014}.

It turns out that there exists a distribution in the literature with the same probability mass function. This is called the \emph{beta-binomial distribution}, denoted $\textrm{BetaBin}(n,\alpha,\beta)$, which depends on the size $n$, and two positive real parameters $\alpha$ and $\beta$. A random variable having this distribution counts the number of successes in $n$ Bernoulli trials, where the success probability is not fixed but is drawn from the well-known \emph{beta distribution}~\cite[Section 6.2.2]{johnson-et-al-2005}.
Recall that the beta distribution $B(\alpha, \beta)$ is a continuous distribution on $[0, 1]$
with probability density proportional to $x^{\alpha - 1} (1-x)^{\beta - 1}$. The normalizing
constant is the beta function
\[
B(x,y) = \frac{\Gamma(x)\Gamma(y)}{\Gamma(x+y)},
\]
where $\Gamma$ is the gamma function
\[
\Gamma(x) = \int\limits_{0}^{\infty} z^{x-1} e^{-z} dz.
\]
The probability that a $\textrm{BetaBin}(n,\alpha,\beta)$ random variable $X$ takes value $x$ is given explicitly by
\begin{equation}
\label{onept corr}
\mathbb{P}(X = x) = \binom{n}{x} \frac{B(x + \alpha, n - x + \beta)}{B(\alpha, \beta)}.
\end{equation}
When $\alpha$ and $\beta$ are positive integers, the beta-binomial distribution becomes the \emph{negative hypergeometric distribution}.

Starting from \eqref{s.s. prob eta_i = alpha t=1} and writing $\rf^{\overline{\alpha}} = \Gamma(\rf + \alpha)/\Gamma(\rf)$, a short calculation shows that
\begin{equation}
\label{s.s. prob eta_i = alpha t=1 betabinom}
    \avg{\eta_i= \alpha} =  \binom{n}{\alpha}
    \frac{B(\alpha+\rf,n-\alpha+(L-1)\rf)}{B(\rf,(L-1)\rf)}.
\end{equation}
Therefore, the distribution of particles at any site is given by the beta-binomial distribution 
$\textrm{BetaBin}(n, \alpha = \rf, \beta = (L-1)\rf)$.

From standard facts about the beta-binomial distribution~\cite[Equation (6.12)]{johnson-et-al-2005}, the average number of particles per site is
\[
n \frac{\alpha}{\alpha + \beta} = \frac{n}{L},
\]
which is evident from the translation-invariance of the \asip{}, and the variance of the number of particles per site is
\begin{equation}
    \label{eq:theoretical var when t=1}
\frac{n \alpha \beta (\alpha + \beta +n) }{(\alpha + \beta)^2 (\alpha + \beta +1)} =
\frac{n(L-1)(L\rf+n)}{L^2(L\rf+1)} 
\end{equation}
Let $\rho \in [0, 1]$ be a fixed constant.
If we let $L, n \to \infty$ such that $n/L \to \rho$, then the variance approaches $\rho (\rf + \rho)$.
See \cref{tab:one point variance} for a comparison between the theoretical value of the variance and numbers from a numerical simulation for the $(0, 1, \rf)$~ASIP with 6 sites and 15 particles. 

\begin{table}[h!]
\begin{center}
        \begin{tabular}{|c|c|c|c|}
            \hline
            $\theta$ & Theoretical value & Simulation result & Error \\
            \hline
            $1/10 = 0.1$ & 20.3125 & 20.4164 & 0.50 \% \\
            \hline
            $1/7 = 0.143$ & 17.78846 & 17.8761 & 0.48 \% \\
            \hline
            $1/4 = 0.25$ & 13.75 & 14.0828 & 2.40 \% \\
            \hline
            1 & 6.25 & 6.2558 & 0.09 \% \\
            \hline
            3 & 3.61842 & 3.5551 & 1.75 \% \\
            \hline
            6 & 2.87162 & 2.8769 & 0.18 \% \\
            \hline
            10 & 2.56147 & 2.5776 & 0.39 \% \\
            \hline
        \end{tabular}
        \caption{Comparision of the variance of the occupation number at the first site in the $(0, 1, \rf)$~ASIP with $L=6$ and $n=15$. The simulation results are averaged over 10000 steady states for 20 different runs.}
        \label{tab:one point variance}
\end{center}
\end{table}

\subsection{Current at $t = 1$}
Just as for the one-point correlation, we do not have a closed-form formula for the current for general values of $t$ due to the lack of a closed-form expression for the partition function. So we will only calculate the current at $t = 1$. This has been computed in the grand canonical ensemble in \cite[Page 526]{cao-etal-2014}.

We derive the current by analyzing the net particle flow between neighbouring sites. Let us consider two consecutive sites labelled $i$ and $i+1$ having $\alpha$ and $\beta$ particles respectively. From the rates described in \eqref{forward rate t=1} and \eqref{backward rate t=1},
we can write down the net rate of flow of particles between the two sites as 
\[
\alpha(\rf+\beta) -q \beta(\rf+\alpha) = (1-q)\alpha \beta + (\alpha -q \beta) \rf.
\]
The steady state current $J$ is obtained by averaging this net flow over the steady state between these two sites where
\begin{equation}
\label{curr-formula}
    J = \sum_{\alpha=0}^{n} \sum_{\beta =0}^{n-\alpha} \big( 
    (1-q)\alpha \beta + (\alpha -q \beta) \rf \big) 
    \avg{\eta_i = \alpha,\eta_{i+1} = \beta}.
\end{equation}

The two-point correlations also follow from the product form of the steady state weights as we showed for one-point correlations in \eqref{s.s. prob eta_i = alpha t=1}, and we obtain 
\begin{equation}
    \avg{\eta_i = \alpha, \eta_{i+1}=\beta} = \binom{n}{\alpha} \binom{n - \alpha}{\beta}
    \rf^{\overline{\alpha}} \rf^{\overline{\beta}} \frac{((L-2)\rf)^{\overline{n-\alpha-\beta}}}{(L\rf)^{\overline{n}}}.
\end{equation}
We compute the current by evaluating each term in the sum separately. The second term in the summation formula for $J$ in \eqref{curr-formula} is 
\begin{equation}
    \rf \sum_{\alpha=0}^{n} \sum_{\beta =0}^{n-\alpha} (\alpha -q \beta)  \avg{\eta_i = \alpha,\eta_{i+1} = \beta} = \rf \sum_{\alpha=0}^{n} \sum_{\beta =0}^{n} (\alpha -q \beta)\avg{\eta_i = \alpha,\eta_{i+1} = \beta},
\end{equation}
where we have used the fact that $\avg{\eta_i = \alpha,\eta_{i+1} = \beta} = 0$ whenever $\alpha+\beta > n$. Expanding this further, we get
\begin{multline}
        \rf \sum_{\alpha=0}^{n} \alpha \Bigl( \sum_{\beta =0}^{n} \avg{\eta_i = \alpha,\eta_{i+1} = \beta} \Bigr) - q \rf \sum_{\beta=0}^{n} \beta \Bigl( \sum_{\alpha =0}^{n} \avg{\eta_i = \alpha,\eta_{i+1}  = \beta} \Bigr) \\ =\rf \sum_{\alpha=0}^{n} \alpha \avg{\eta_i = \alpha} - q \rf \sum_{\beta=0}^{n} \beta \avg{\eta_{i+1} = \beta},
\end{multline}
which is just the difference of two one-point correlations. Thus
\begin{equation}
\label{current calculation first term answer}
    \rf \sum_{\alpha=0}^{n} \sum_{\beta =0}^{n-\alpha} (\alpha -q \beta)  \avg{\eta_i = \alpha,\eta_{i+1} = \beta} = \frac{n\rf}{L} - q\frac{n\rf}{L} = \frac{(1-q)n\rf}{L}.
\end{equation}
The evaluation of the first term for $J$ in \eqref{curr-formula} is slightly more complicated. We need to determine 
\[
\sum_{\alpha=0}^{n} \sum_{\beta =0}^{n-\alpha} \alpha \beta \, \avg{\eta_i=\alpha,\eta_{i+1}=\beta}.
\]
Using \eqref{wt_eta1}, the two-point correlations can be easily derived, and this can be written as
\begin{equation}\label{current calculation second term}
\sum_{\alpha=0}^{n} \sum_{\beta =0}^{n-\alpha} \alpha \beta \binom{n}{\alpha} \binom{n-\alpha}{\beta}
\frac{\rf^{\overline{\alpha}} \rf^{\overline{\beta}} ((L-2)\rf)^{\overline{n - \alpha - \beta}}}
{ (L\rf)^{\overline{n}}}.
\end{equation}
Letting $\kappa = (L-2)\rf$, the right hand side above becomes
\[
\frac{n!}{ (L\rf)^{\overline{n}}} \sum_{\alpha=1}^{n-1} \sum_{\beta =1}^{n-\alpha} 
\frac{\theta^{\overline{\alpha}}}{(\alpha-1)!}
\frac{\theta^{\overline{\beta}}}{(\beta-1)!}
\frac{\kappa^{\overline{n - \alpha - \beta}}}{(n - \alpha - \beta)!}.
\]
It is clear that the rising factorial in \eqref{rising_factorial} satisfies the identity $x^{\overline{n}} = x (x+1)^{\overline{n-1}}$. 
{Plugging this for $\theta^{\overline{\alpha}}$ and $\theta^{\overline{\beta}}$
using  the multinomial Chu-Vandermonde identity in \eqref{multi-chu-van}, 
we obtain}
\[
n(n-1) \theta^2 \frac{(2\theta +\kappa +2)^{\overline{n-2}}} {(L\rf)^{\overline{n}}}.
\]
Substituting $\kappa$ back and simplifying, we obtain
\begin{equation}
\label{current calculation second term answer}
\frac{n(n-1)\rf}{(L\rf+1)L} .
\end{equation}
Adding the contributions from \eqref{current calculation first term answer} and $(1-q)$ times the expression in \eqref{current calculation second term answer} yields the exact expression for the steady state current,
\begin{equation}
\label{current total final answer}
    J =  \frac{(1-q) \, \rf \, n(L\rf+n)}{L(L\rf+1)}.
\end{equation}
If we let $L, n \to \infty$ such that $n/L \to \rho$, then the current becomes $J = (1-q) \rho(\rf + \rho)$, which is also derived in \cite[Equation (24)]{cao-etal-2014}, and which is $(1-q)$ times the limiting variance.

\section{Phase diagram}
\label{sec:phase diagram}
We analyze the two-dimensional phase diagram of the \asip{} in terms of the parameters $t$ and $\rf$ { which will be fixed constants, in the limit where both $L$ and $n$ are going to infinity}. Although there are no phase transitions, we show that there are crossovers and the steady state looks very different in different regions of the phase diagram.

We understand the most probable configurations 
by examining the polynomial structure of the steady state weights \eqref{wt_eta1} and identifying the dominant terms in various asymptotic limits. We first establish key properties of the polynomial weights. Define the \emph{degree} (resp. \emph{order}) of a polynomial $p$ to be the largest (resp. smallest) exponent for variable $x$ with nonzero coefficient, denoted $\deg_x(p)$ (resp. $\ord_x(p)$). 
One can easily show
\begin{equation}\label{eq: order t weight in theta}
    \ord_t(\wt(\eta)) = 0,
\end{equation}
and
\begin{equation}\label{eq: degree theta weight in theta}
    \deg_{\rf}(\wt(\eta)) = n.
\end{equation}

We now prove
\begin{equation}\label{eq: order theta weight in theta}
    \ord_\rf(\wt(\eta)) = L-n_0(\eta),
\end{equation}
and
\begin{equation}\label{eq: degree t weight in theta}
    \deg_t(\wt(\eta)) = \frac{n(n-3)}{2} + L-n_0(\eta),
\end{equation}
where $n_0(\eta)$ is the number of empty sites in the configuration $\eta$.
From \eqref{eq:defining box bar t}, we see that
\[
\ord_\rf(\bbt{\rf}{m}{t}) =  
\begin{cases}
    0 & m=0,\\
    1 & \textrm{otherwise},
\end{cases}
\]
and, since the $t$-multinomial coefficient has no dependence on $\rf$,
\begin{equation}\label{order in theta}
    \ord_\rf(\wt(\eta)) = \ord_{\rf} \left(\prod\limits_{i=1}^{L}\bbt{\rf}{\eta_i}{t} \right)=L-n_0(\eta).
\end{equation}
Using $\deg_t(\bt{n}!) = n(n-1)/2$ we get
\begin{equation}\label{degree of multinomial}
    \deg_t \left( \qbinomt n{\eta_1,\ldots,\eta_L} \right) 
    =  \deg_t(\bt{n}!) - \sum_{i=1}^L \deg_t(\bt{\eta_i}!)
    = \frac{n^2}{2} - \sum\limits_{i=1}^{L}\frac{\eta_i^2}{2}.
\end{equation}
Now for the remaining factor of $\prod\limits_{i=1}^{L}\bbt{\rf}{\eta_i}{t}$, since
\[
\deg_t(\rf+\bt{j})=  
\begin{cases}
    0 &\textrm{when } j=0,\\
    j-1 &\textrm{otherwise},
\end{cases}
\]
we get
\[
\deg_t(\bbt{\rf}{m}{t}) =  
\begin{cases}
    0 &\textrm{when } m=0,\\
    \frac{(m-1)(m-2)}{2} &\textrm{otherwise},
\end{cases}
\]
and therefore
\begin{equation}\label{degree t2}
\deg_t \left( \prod\limits_{i=1}^{L}\bbt{\rf}{\eta_i}{t} \right) 
= \sum\limits_{\substack{i=1\\\eta_i \neq 0}}^{L}\Bigl( \frac{\eta_i^2}{2} - \frac{3\eta_i}{2} + 1 \Bigr) =\sum\limits_{i=1}^{L}\frac{\eta_i^2}{2} - \frac{3n}{2} + L - n_0(\eta).
\end{equation}
Therefore, using \eqref{wt_eta1} and adding the results from \eqref{degree of multinomial} and \eqref{degree t2}, we get \eqref{eq: degree t weight in theta}.
The reader can verify \eqref{eq: order t weight in theta}-\eqref{eq: degree t weight in theta} for the \asip{} with $L=3$ and $n=4$ by looking at \eqref{eg:L=3,n=4,t factorized}.

We now analyse the phase diagram in various regions. To do so we will now estimate $\bbt{\rf}{\eta}{t}$ and $\qbinomt{n}{\eta_1,\ldots,\eta_L}$ in different ranges of t. We start by taking extreme limits for $t$ and see what simplifications can be made when it is either very small or very large. When $t \ll 1$ we can write
\[
\bt{m} = \sum_{i=0}^{m-1} t^i   
\begin{cases}
    = 0 \qquad &\textrm{when } {m}=0,\\
     \approx 1 \qquad &\textrm{otherwise }.
\end{cases}
\]
This simplifies the $t$-multinomial coefficient to
\begin{equation}\label{eq: small t qbinom simplification}
    \qbinomt{n}{\eta_1,\eta_2,\ldots,\eta_L} \approx 1,
\end{equation}
and $\bbt{\rf}{\eta_i}{t}$ to
\begin{equation}\label{eq: small t bbt eta_i}
    \bbt{\rf}{\eta_i}{t}=(\rf)(\rf+\bt{1})\ldots(\rf+\bt{\eta_i-1})
\begin{cases}
    =1 \qquad &\textrm{when } \eta_i=0,\\
     \approx \rf(\rf+1)^{\eta_i -1} \qquad &\textrm{otherwise}.
\end{cases}
\end{equation}
On the other hand, when $t \gg 1$, the only simplification we have is
\begin{equation}\label{eq: box m large t}
    \bt{m} = \sum_{i=1}^{m-1} t^{{i}} 
\begin{cases}
    = 1 \qquad &\textrm{when } m=0,\\
 \approx t^{m-1} \qquad &\textrm{otherwise, }
\end{cases}
\end{equation}

In the following subsections, we will analyze the cases when $t \ll 1$, $t = 1$ and $t \gg 1$ in that order. Finally, we will consider the special case $\rf t = 1$.

\subsection{$t \ll 1$ and $\rf \ll 1$}
\label{sec:t<1 rf<1}
In this regime, we see strong particle aggregation phenomena. The condition $\rf \ll 1$ in \eqref{eq: small t bbt eta_i} simplifies it to
\begin{equation}\label{eq: small t small theta bbt eta_i}
    \bbt{\rf}{\eta_i}{t}
\begin{cases}
    =1 \qquad &\textrm{when } \eta_i=0,\\
     \approx \rf\qquad &\textrm{otherwise},
\end{cases}
\end{equation}
and using \eqref{eq: small t qbinom simplification} along with it we get
\begin{equation}
    \wt(\eta) \approx \rf^{L-n_0(\eta)}.
\end{equation}
Since $\rf \ll 1$, the configurations with the highest probability will be the ones which minimize $L-n_0(\eta)$ i.e. maximizing $n_0(\eta)$. In any given configuration $\eta$ the maximum number of empty sites possible are
\begin{equation}\label{max n_0}
    \textrm{max}(n_0(\eta)) = L-1,
\end{equation}
which happens when all the particles occupy the same site. Thus, we see {grouping} of particles in a single site, showing the phenomenon of \textit{strong condensation}, first 
coined in \cite{evans-waclaw-2014}.
Calculations for the example in \eqref{eg:L=3,n=4,t factorized} with the parameter values $\rf=t=0.0001$ gives the steady state probabilities
\begin{multline*}
\pi(0,0,4) = 0.33,
\pi(0,1,3)  = 3.32\times 10^{-5}, \\
\pi(0,2,2) = 3.33 \times 10^{-5},
\pi(1,1,2) = 3.33 \times 10^{-9}.
\end{multline*}

\begin{figure}[h!]
    \centering
    \includegraphics[width=0.45\linewidth]{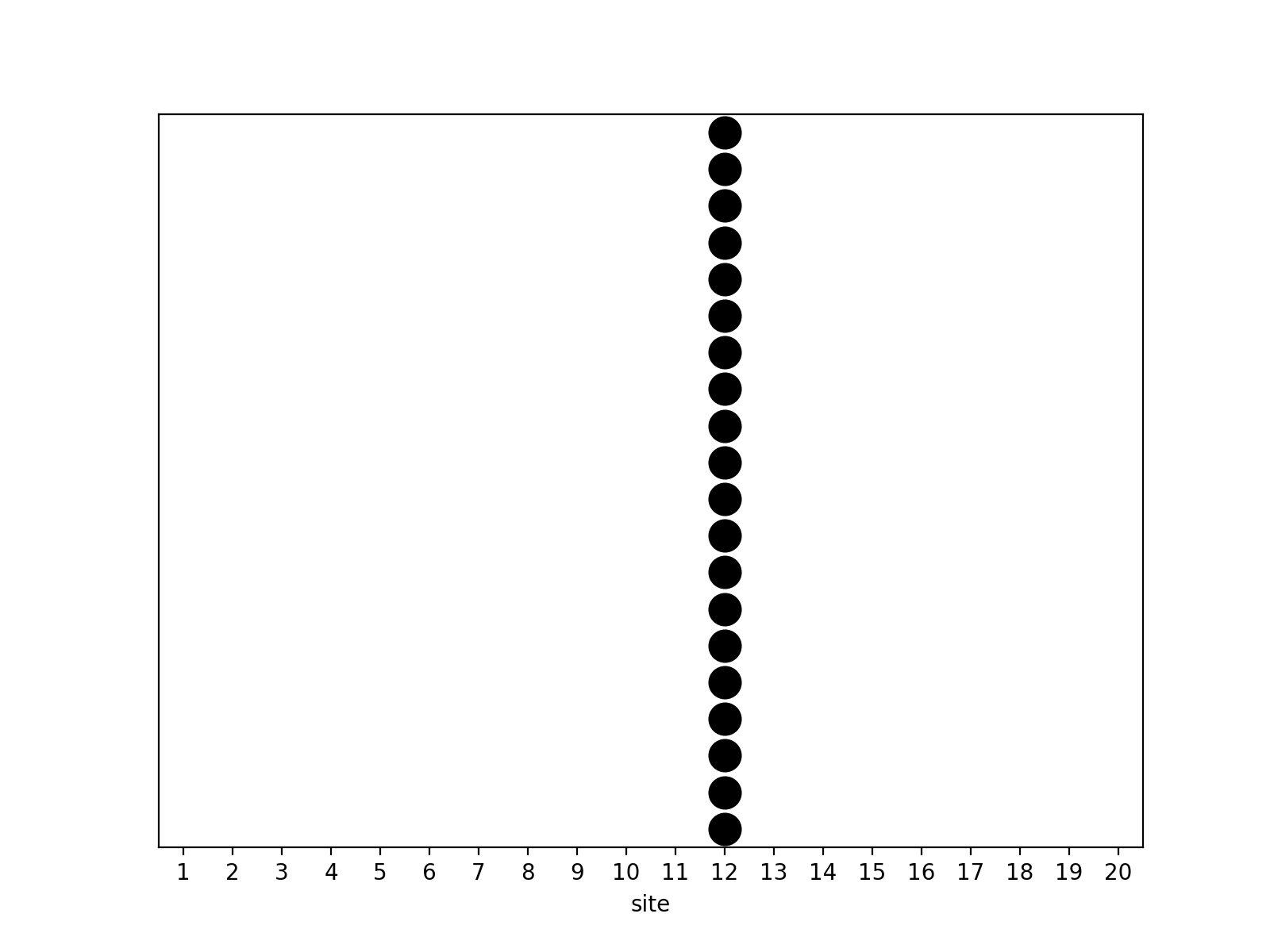}
    \caption{A snapshot of the $(0, 0.0001, 0.0001)$~ASIP with $L=20$ and $n=20$ in steady state.}
    \label{fig:5.1 figs}
\end{figure}

See the movie \texttt{t\_.0001\_theta\_.0001.mp4} among the ancillary files for a simulation of the $(0, 0.0001, 0.0001)$~ASIP with $L = n = 20$. A snapshot from that movie is shown in \cref{fig:5.1 figs}.

\subsection{$t \ll 1$ and $\rf \gg 1$}\label{sec: approximately uniform}
In this regime, we see an approximate uniform distribution across configurations.
Using \eqref{eq: small t qbinom simplification} and adding the condition $\rf \gg 1$ in \eqref{eq: small t bbt eta_i}, we obtain
\begin{equation}\label{eq: small t big theta bbt eta_i}
    \bbt{\rf}{\eta_i}{t}
\begin{cases}
    =1 \qquad &\textrm{when } \eta_i=0,\\
     \approx \rf^{\eta_i} \qquad &\textrm{otherwise}.
\end{cases}
\end{equation}
Therefore for any state $\eta$ we will have
\begin{equation}
    \wt(\eta) \approx \prod\limits_{i=1}^L \rf(\rf+1)^{\eta_i-1} \approx \rf^n,
\end{equation}
which is roughly the same for all configurations, so
\begin{equation}
    \pi(\eta) \approx \frac{1}{\ds\binom{L+n-1}{n}}.
\end{equation}

Consider states where the particles are as evenly spread out as possible. Such states will have a higher probability as compared to others due to the higher value of the corresponding multinomial coefficient. One might expect to see {these in simulations because they have the highest probability. However, that is not the case because of entropic considerations.}

\begin{figure}[h!]
    \centering
    \includegraphics[width=0.45\linewidth]{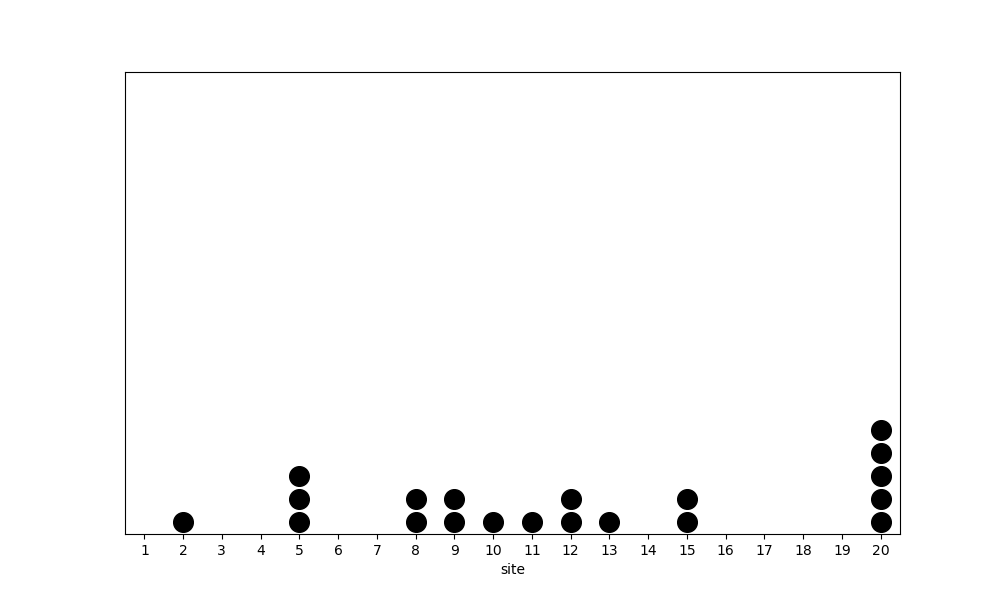}
    \caption{A snapshot of the $(0, 0.0001, 500)$~ASIP with $L=20$ and $n=20$ in steady state.}
    \label{fig:5.2 figs}
\end{figure}

Calculations for the example in \eqref{eg:L=3,n=4,t factorized} with the parameter values $\rf=1000$ and $t=0.01$ give the steady state probabilities
\begin{multline*}
\pi(0,0,4) \approx 6.60 \times 10^{-2},
\pi(0,1,3) \approx 6.66 \times 10^{-2}, \\
\pi(0,2,2)  \approx 6.66 \times 10^{-2},
\pi(1,1,2) \approx 6.72 \times 10^{-2}.
\end{multline*}
See the movie \texttt{t\_.0001\_theta\_500.mp4} among the ancillary files for a simulation of the $(0, 0.0001, 500)$~ASIP with $L = n = 20$. A snapshot from that movie is shown in \cref{fig:5.2 figs}.

\subsection{$t = 1$}
\label{sec:t=1}

{This case has already been studied before~\cite{grosskinsky-etal-2011,Jatuviriyapornchai2020}. 
Our calculations and simulations match their conclusion 
suggesting that condensation occurs for $\rf \ll 1$
and hyperuniformity occurs for $\rf \gg 1 $.
We do not comment on this any further.
}

\subsection{$\rf \ll 1 \ll t$ and $\rf t \ll 1$}\label{subsec: theta t < 1}
\begin{figure}[h!]
    \centering
    \includegraphics[width=0.45\linewidth]{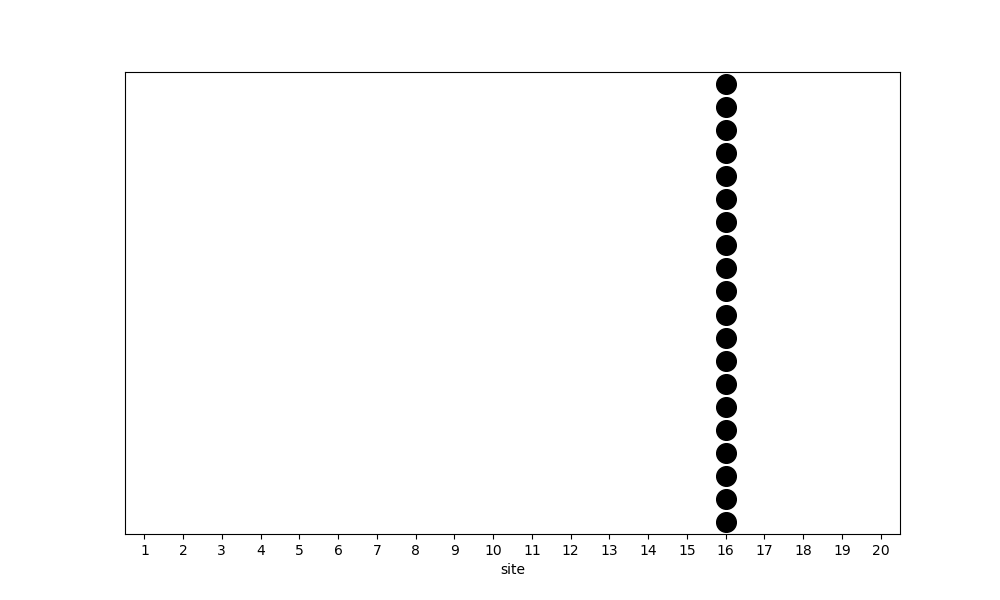}
    \caption{A snapshot of the $(0, 500, 0.0001)$~ASIP with $L=20$ and $n=20$ in steady state.}
    \label{fig:5.5 figs}
\end{figure}

This region again favours stronger condensation similar to \cref{sec:t<1 rf<1}. 
Using \eqref{eq: box m large t} along with $\rf \ll 1$ will simplify \eqref{eq:defining box bar t} as
\begin{equation}
    \bbt{\rf}{\eta_i}{t}
\begin{cases}
    = 1 \qquad &\textrm{when } \eta_i=0,\\
    \approx \rf t^{\frac{(\eta_i-1)(\eta_i-2)}{2}} \qquad &\textrm{otherwise}.
\end{cases}
\end{equation}
Combining this with the largest term from the multinomial coefficient we can write
\[
\wt(\eta) = \rf^{L-n_0(\eta)} t^{\deg_t(\wt(\eta))} +\textrm{ other terms }
\]
where $n_0(\eta)$ is the number of empty sites in the configuration $\eta$. Using the formula for $\deg_t(\wt(\eta))$ given in \eqref{eq: degree t weight in theta} we obtain
\begin{equation}\label{eq: weight t bigger than 1 theta less than 1}
    \wt(\eta) \approx \rf^{L-n_0(\eta)} t^{\frac{n(n-3)}{2}+L-n_0(\eta)} = (\rf t)^{L-n_0(\eta)} t^{\frac{n(n-3)}{2}}.
\end{equation}
The configuration with the highest steady state probability will therefore be dictated by the value of $\rf t$ which we will now explore. To maximize the weight \eqref{eq: weight t bigger than 1 theta less than 1} when $\rf t \ll 1$, we want to minimize $L-n_0(\eta)$, and hence maximize $n_0(\eta)$. So the dominant steady state becomes the condensate states where the value $n_0(\eta)$ takes its maximum value $L-1$ as in \eqref{max n_0}. Calculations for the example in \eqref{eg:L=3,n=4,t factorized} with the parameter values $\rf = 0.000002$ and $t=1000$ give the steady state probabilities
\begin{multline*}
\pi(0,0,4) \approx 0.33,
\pi(0,1,3) \approx 6.62 \times 10^{-4}, \\
\pi(0,2,2)  \approx 6.62 \times 10^{-4},
\pi(1,1,2) \approx 1.33 \times 10^{-6}.
\end{multline*}
See the movie {\texttt{t\_500\_theta\_.0001.mp4}} among the ancillary files for a simulation of the $(0, 500, 0.0001)$~ASIP with $L = n = 20$. A snapshot from that movie is shown in \cref{fig:5.5 figs}.

\subsection{$\rf \ll 1 \ll t$ and $\rf t \gg 1$ }\label{subsec: introducing flattening}
\begin{figure}[h!]
    \centering
    \includegraphics[width=0.45\linewidth]{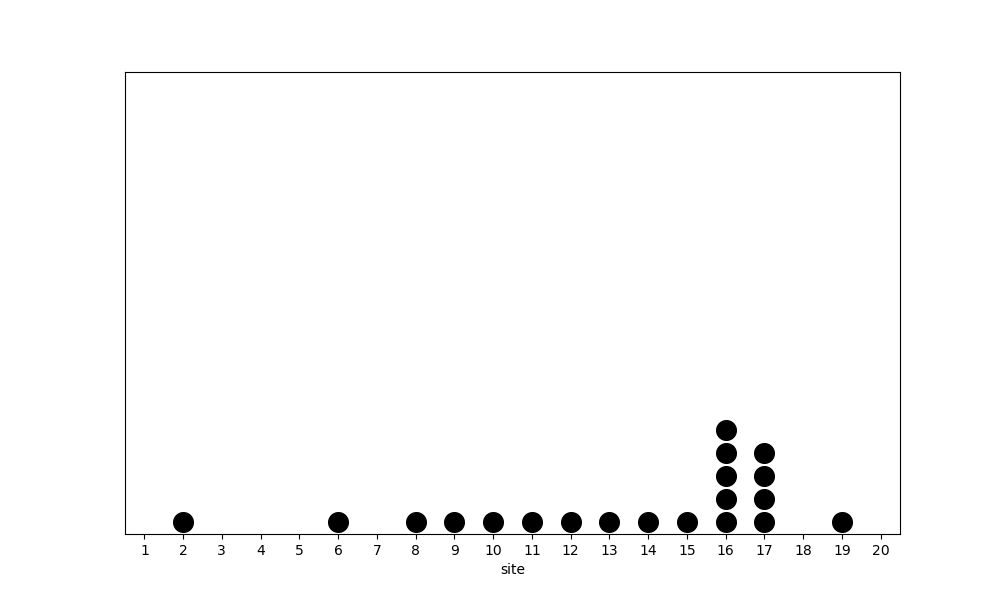}
    \caption{{A snapshot of the $(0, 1000, 0.1)$~ASIP with $L=20$ and $n=20$ in steady state}}
    \label{fig:5.6 figs}
\end{figure}

For $\rf \ll 1 \ll t$ we analyze \eqref{eq: weight t bigger than 1 theta less than 1}, this time for $\rf t \gg 1$. 
Here $L-n_0(\eta)$ must be maximized. 
For any state $\eta$ we have
\[
\min(n_0(\eta)) = 
\begin{cases}
    L-n & \textrm{when } n <L,\\
    0 & \textrm{otherwise}.
\end{cases}
\]
Therefore, when $n \leq L$, there will at most be one particle per site in the likely configurations. When $L \leq n$, all sites are occupied. This time, hyperuniform states will dominate because the $t$-multinomial coefficient is larger for such states. 
Numerical calculations confirm the preference for configurations with few particles per site. 
{As mentioned earlier, this system seems to demonstrate hyperuniformity similar to earlier studies~\cite{chleboun-et-al-2018}.}
For example,in \eqref{eg:L=3,n=4,t factorized} with parameters $\rf = 0.002$ and $t=10^6$, we get the steady state probability distribution
\begin{multline*}
\pi(0,0,4) \approx 8.32 \times 10^{-8},
\pi(0,1,3) \approx 1.66 \times 10^{-4}, \\
\pi(0,2,2) \approx 1.66 \times 10^{-4},
\pi(1,1,2)  \approx 0.33.
\end{multline*}
See the movie {\texttt{t\_1000\_theta\_.1.mp4}} among the ancillary files for a simulation of the {$(0, 1000, \linebreak 0.1)$~ASIP} with $L = n = 20$. A snapshot from that movie is shown in \cref{fig:5.6 figs}.

\subsection{$1 \ll \rf \ll t$}\label{sec:1<rf<t}

Using \eqref{eq: box m large t} and simplifying
\begin{equation}
    \bbt{\rf}{\eta_i}{t}
\begin{cases}
    = 1 \qquad &\textrm{when } \eta_i=0,\\
    \approx \rf, &\textrm{for }\eta_i=1,\\
    \approx \rf^2 t^{\frac{(\eta_i-1)(\eta_i-2)}{2}} \qquad &\textrm{otherwise},
\end{cases}
\end{equation}
for $\rf \gg 1$ gives the steady state weights
\begin{equation}
    \wt(\eta) \approx \rf ^{n_1(\eta)} (\rf^2)^{L-n_1(\eta)-n_0(\eta)} t^{\deg_t(\wt(\eta))} = \rf^{L-n_1(\eta)-n_0(\eta)} (\rf t)^{L-n_0(\eta)} t^{\frac{n(n-3)}{2}}.
\end{equation}
Since $\rf \ll \rf t$ and the power of $t$ is independent of $\eta$, we look to maximize $L-n_0(\eta)$. This is possible when $n_0(\eta)=0$ i.e. all the sites are filled. Calculations for the example in \eqref{eg:L=3,n=4,t factorized} with the parameter values $\rf = 100$ and $t=10000$ give the steady state probabilities
\begin{multline*}
\pi(0,0,4) \approx 3.36 \times 10^{-13},
\pi(0,1,3) \approx 3.36 \times 10^{-7}, \\
\pi(0,2,2)  \approx 3.36 \times 10^{-5},
\pi(1,1,2) \approx 0.33.
\end{multline*}

\begin{figure}[h!]
    \centering
    \includegraphics[width=0.45\textwidth]{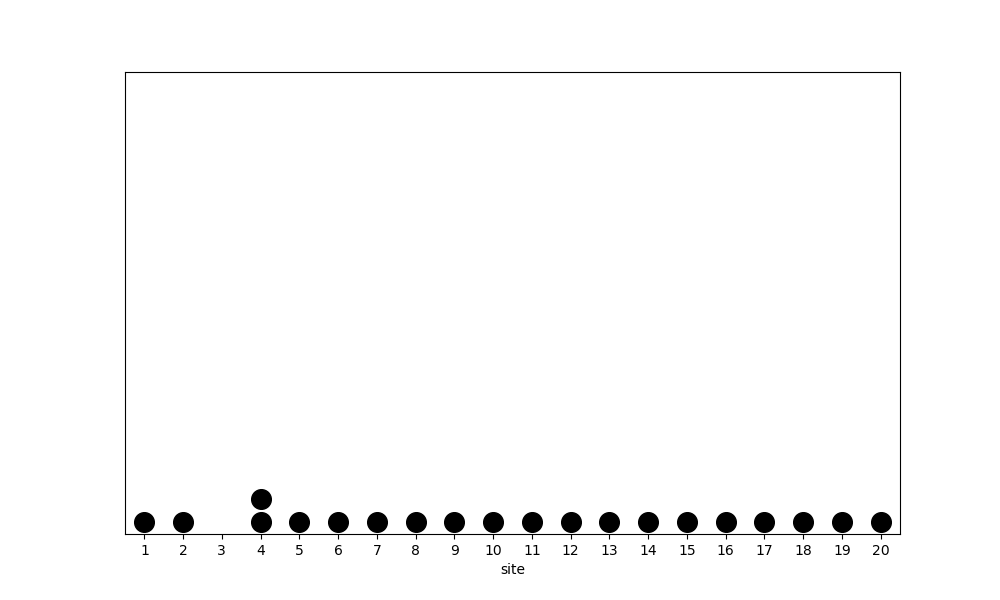}
    \caption{{A snapshot of the $(0, 10000, 100)$~ASIP with $L=20$ and $n=20$ in steady state}}
    \label{fig:5.7 figs}
\end{figure}

See the movie {\texttt{t\_10000\_theta\_100.mp4}} among the ancillary files for a simulation of the $(0, 10000, 100)$~ASIP with $L = n = 20$. A snapshot from that movie is shown in \cref{fig:5.7 figs}.

\subsection{$1 \ll t  \ll t^n \ll \rf$}\label{sec:1<t<t^n<rf}
\begin{figure}[h!]
    \centering
    \includegraphics[width=0.45\linewidth]{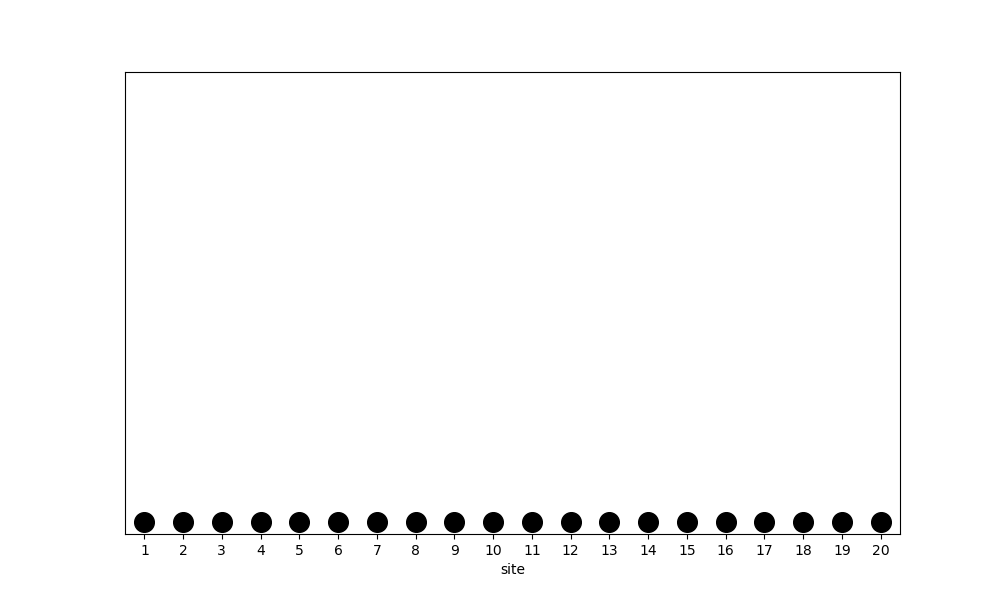}
    \caption{A snapshot of the $(0, 100, 10^{50})$~ASIP with $L=20$ and $n=20$ in steady state.}
    \label{fig:5.8 figs}
\end{figure}

Using \eqref{eq: box m large t} and simplifying 
\begin{equation}
    \bbt{\rf}{\eta_i}{t} \approx \rf^{\eta_i},
\end{equation}
we can write, after dropping the common factor of $\rf^n$ from all the weight expressions,
\begin{equation}
    \wt(\eta) \approx \qbinomt{n}{\eta_1,\ldots,\eta_L}.
\end{equation}
This $t$-multinomial coefficient is greatest when the particles will be maximally spread out, i.e., in the configuration which was discussed in \cref{subsec: introducing flattening}. Calculations for the example in \eqref{eg:L=3,n=4,t factorized} with the parameter values $\rf = 10^{50}$ and $t=10000$ give the steady state probabilities
\begin{multline*}
\pi(0,0,4) \approx 3.23 \times 10^{-11},
\pi(0,1,3) \approx 3.26 \times 10^{-5}, \\
\pi(0,2,2)  \approx 3.26 \times 10^{-3},
\pi(1,1,2) \approx 0.33.
\end{multline*}
See the movie \texttt{t\_100\_theta\_1e+50.mp4} among the ancillary files for a simulation of the $(0, 100, 10^{50})$~ASIP with $L = n = 20$. A snapshot from that movie is shown in \cref{fig:5.8 figs}.

\subsection{$\rf t = 1$}

\begin{figure}[h!]
     \centering
     \begin{subfigure}[b]{0.45\textwidth}
         \centering
         \includegraphics[width=\textwidth]{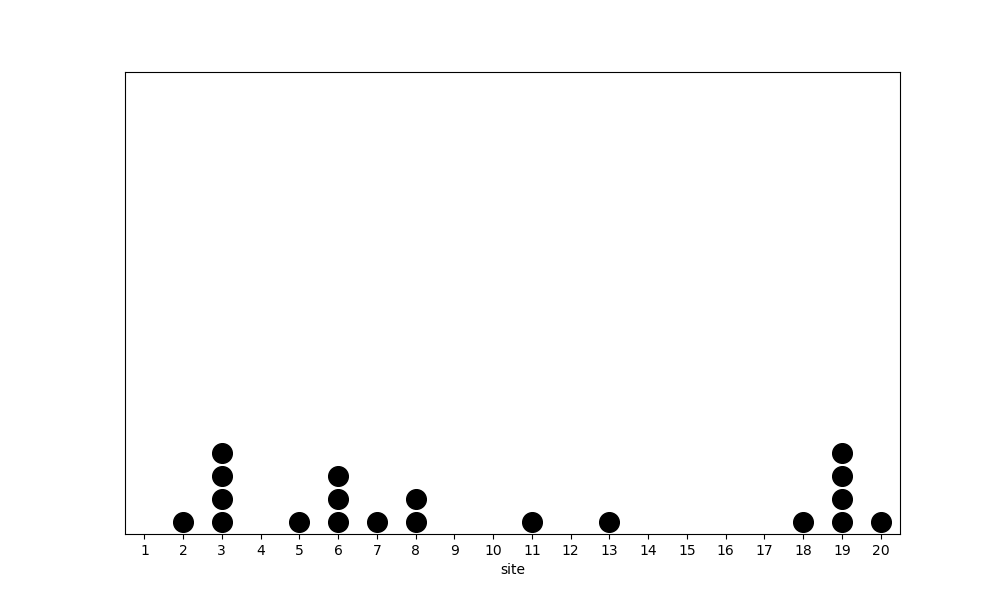}
         \caption{}
         \label{fig:5.9a_a}
     \end{subfigure}
     \begin{subfigure}[b]{0.45\textwidth}
         \centering
         \includegraphics[width=\textwidth]{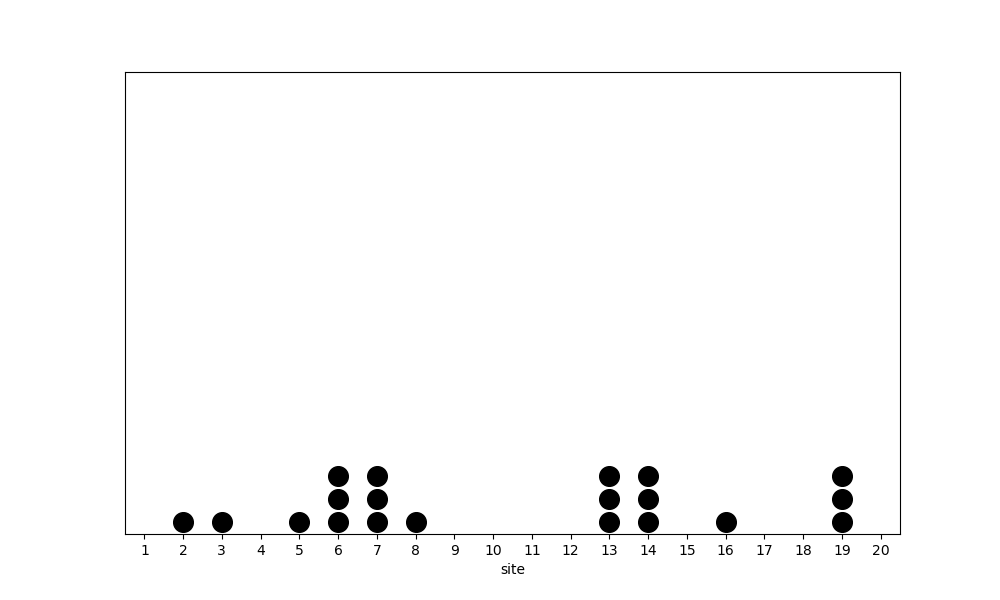}
         \caption{}
         \label{fig:5.9a_b}
         
     \end{subfigure}

     \begin{subfigure}[b]{0.45\textwidth}
         \centering
         \includegraphics[width=\linewidth]{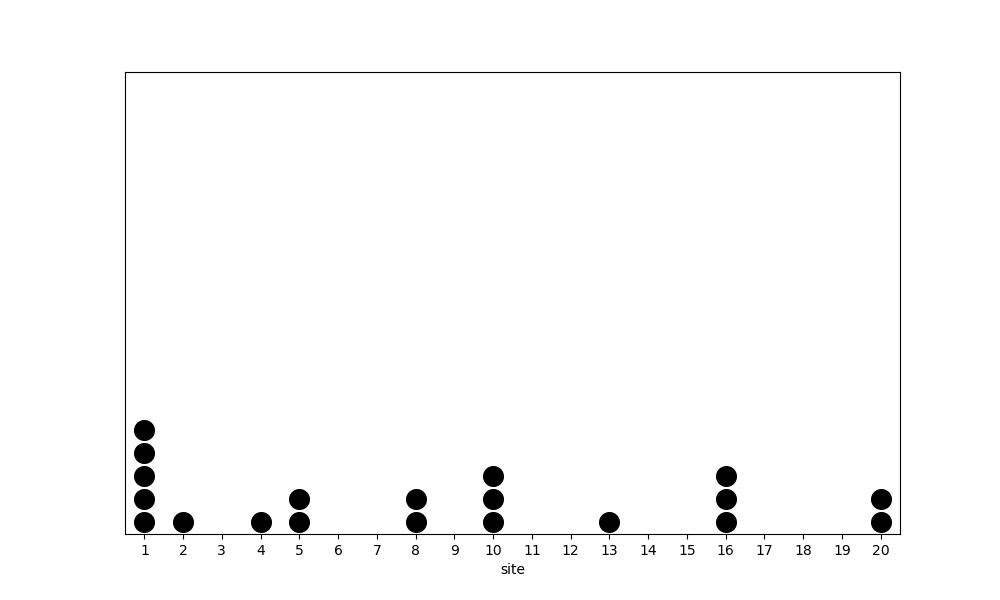}
         \caption{}
         \label{fig:5.9a_c}
         
     \end{subfigure}
     
        \caption{Snapshots for $L=20$ and $n=20$ in steady state for (a) $(0, 0.01, 100)$~ASIP, (b) $(0, 1, 1)$~ASIP and {(c)} {$(0, 10, 0.1)$~ASIP}}
        \label{fig:5.9a figs}
\end{figure}

As calculated before in \eqref{eq: uniform when theta t = 1}, we get a uniform distribution. 
Simulations in \texttt{t\_.01\_theta} \texttt{\_100.mp4}, \texttt{t\_1\_theta\_1.mp4} 
and {\texttt{t\_10\_theta\_.1.mp4}} are added in the ancillary files for three different values of $t$ and $\rf$ where the product is $1$.
Snapshots of these movies are shown in \cref{fig:5.9a_a,fig:5.9a_b,fig:5.9a_c} respectively.

\begin{figure}[htbp!]
    \centering
    \includegraphics[width=0.8\linewidth]{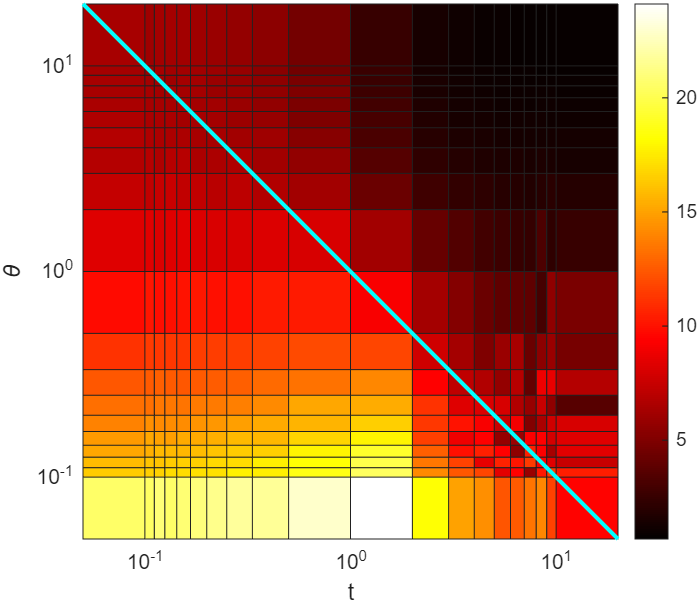}
    \caption{Variance of the single site occupation in for the \asip{} with $L=6$ and $n=15$ { plotted as a heat map on a logarithmic scale. The diagonal line corresponds to $\rf = 1/t$.
    The variance seems to go to zero on the top right, suggesting hyperuniform behaviour.}
        The parameters $t,\rf$ each range over the 21 values $\{n,1/n \mid n \in \{1, \dots, 10\}\} \cup \{20, 1/20\}$. The variance is computed by averaging over 10000 steady state configurations for 20 different runs. }
        \label{fig:variance 3d plot}
\end{figure}

The $\rf t=1$ curve marks a smooth crossover from the strong condensation observed when 
$\rf t \ll 1$ in \cref{sec:t<1 rf<1,subsec: theta t < 1}  
to {hyperuniformity} observed when $\rf t  \gg 1 $ in \cref{subsec: introducing flattening,sec:1<rf<t,sec:1<t<t^n<rf}.
This matches the trend observed in the variance plot shown in \cref{fig:variance 3d plot}.

\section{Palindromicity and antipalindromicity of the weights} 
\label{sec:palindromicity and antipalindromicity}

As mentioned above, we will use the equivalent formulation of the rates in \eqref{forward rate a and t} and \eqref{backward rate a and t} when $t \neq 1$.
Using the forward rate \eqref{forward rate a and t} in \eqref{eq: evans eqn 1} for the calculation of the steady state weights, we get
\[
f_a(m) = \prod_{i = 1}^{m} \frac{u(1, i-1)}{u(i, 0)} = \prod_{i = 1}^{m} \frac{1-\pa t^{i-1}}{(1-\pa) \bt{i}}.
\]
Recall the \emph{$t$-Pochhammer symbol} given by
\begin{equation}
    (\pa;t)_{m} = \prod^{m-1}_{k=0} (1-\pa t^{k}) = (1-\pa)(1-\pa t)
    \dots (1-\pa t^{m-1}).
\end{equation}
Then we obtain, up to overall normalisation,
\begin{equation}
\label{density2}
f_a(m) = \frac{(\pa;t)_{m}}{(t; t)_{m}},
\end{equation}
using $(t;t)_m = [m]_t! (1-t)^m$.
We thus obtain the steady state weight
\begin{equation}
\label{wt_eta2}
\wta(\eta) = \prod_{i=1}^L f_a(\eta_i) = 
(t; t)_n \prod_{i=1}^L \frac{(\pa;t)_{\eta_i}}{(t; t)_{\eta_i}}
\end{equation}
after rescaling. 
This makes all the weights polynomials in $\pa$ and $t$. 

For example, the steady weights of the \asip{} with $L=3$ and $n=4$ in \eqref{eg:L=3,n=4,t factorized}, when rewritten in these variables, are
\begin{equation}
\label{wt example 2}
\begin{aligned}
\wta(0,0,4) =& \;(1-\pa)(1-\pa t)(1-\pa t^2)(1-\pa t^3), \\
    \wta(0,1,3) = & \; (1 + t) (1 + t^2)(1-\pa)^2 (1-\pa t)(1-\pa t^2),\\
    \wta(0,2,2) =& \;(1 + t^2) (1 + t + t^2) ( 1 - \pa)^2 (1 - \pa t)^2, \\
    \wta(1,1,2) =& \; (1 + t) (1 + t^2) (1 + t + t^2)(1 - \pa)^3 (1-\pa t).
\end{aligned}
\end{equation}

We denote the partition function in these variables as
\begin{equation}
\label{pf sum 2}
\Za_{L,n} = \sum_{\eta \in \Omega_{L,n}} \wta(\eta).
\end{equation}
Although the steady state weights factor, we do not seem to obtain simple formulas for the ordinary or exponential generating functions of $\Za_{L,n}$. There is another variant, known as the \emph{Eulerian generating function}~\cite{goldman-rota-1970} that will be useful to us. We will now show that the Eulerian generating function of the partition function is given by the product formula,
\begin{equation}
\label{pf-gf}
\sum_{n = 0}^\infty \Za_{L,n} \frac{x^n }{(t; t)_n} = 
\left( \frac{(ax; t)_\infty}{(x;t)_\infty} \right)^L.
\end{equation}

To begin, we see that the generating function for $f_a(m)$ is
\begin{equation}
\label{eq:inf sum q binom}
\sum_{m=0}^{\infty} f_a(m) z^m = 
\sum_{m=0}^{\infty} \frac{(a;t)_m}{(t; t)_m} z^m = \frac{(az; t)_\infty}{(z;t)_\infty},
\end{equation}
using the general formulation of the $q$-binomial theorem~\cite[Equation (1.3.2)]{gasper-rahman-2004}.
The Eulerian generating function of the partition function is then given,
using \eqref{wt_eta2}, by
\begin{equation}
\sum^\infty_{n=0} \Za_{L,n} \frac{x^n }{(t; t)_n}
= 
\sum_{n=0}^{\infty} \left( \sum_{\eta \in \Omega_{L,n}} 
(t; t)_n \prod_{i=1}^L \frac{(\pa;t)_{\eta_i}}{(t; t)_{\eta_i}}
 \right) \frac{x^n}{(t; t)_n}.
\end{equation}
The factor of $(t; t)_n$ cancels and we can split the product terms to get
\begin{equation}
\sum_{\eta_1 \geq 0} \sum_{\eta_2 \geq 0} \cdots \sum_{\eta_L \geq 0} 
\frac{(a;t)_{\eta_1}}{(t; t)_{\eta_1}} 
\frac{(a;t)_{\eta_2}}{(t; t)_{\eta_2}} \cdots 
\frac{(a;t)_{\eta_L}}{(t; t)_{\eta_L}} x^{\eta_1} x^{\eta_2} \cdots x^{\eta_L}.
\end{equation}
Each of the $L$ sums can be performed independently using \eqref{eq:inf sum q binom} to obtain \eqref{pf-gf}.

We now study the symmetry properties of the steady state weights. Recall the order and degree of a polynomial defined in \cref{sec:phase diagram}. One can easily compute that $\ord_t(\wta(\eta)) = \ord_\pa(\wta(\eta)) = 0$.
We begin by computing the polynomial degrees, which are crucial for establishing the symmetry properties.
First, note that
\[
\deg_a(\pa;t)_{\eta_i} = \eta_i
\]
and
\[
\deg_t(\pa;t)_{\eta_i} = \eta_i(\eta_i-1)/2.
\]
Thus, the degree of the product of these Pochhammer terms is
\begin{equation}
    \deg_a \Bigl(\prod_{i=1}^{L}(\pa;t)_{\eta_i} \Bigr) = \eta_1 + \cdots + \eta_L = n, 
\end{equation}
and
\begin{equation}
    \deg_t \Bigl(\prod_{i=1}^{L}(\pa;t)_{\eta_i} \Bigr) = \frac{\eta_1(\eta_1-1)}{2} + \cdots + \frac{\eta_L(\eta_L-1)}{2} = \frac{\eta_1^2+\cdots+\eta_L^2 - n}{2}.
\end{equation}
Combining these  with \eqref{degree of multinomial} we get the degrees as
\begin{equation}
    \deg_a (\wta(\eta_1,\ldots,\eta_L)) = n,
\end{equation}
and
\begin{equation}
    \deg_t (\wta(\eta_1,\ldots,\eta_L)) = \frac{n^2 - \eta_1^2 - \ldots - \eta_L^2}{2} + \frac{\eta_1^2+\ldots+\eta_L^2 - n}{2} = \frac{n(n-1)}{2}.
\end{equation}
Notice that the orders and degrees of $\wta(\eta)$ in $t$ and $\pa$ are independent of the exact configuration $\eta$ and only depend on $n$, the total number of particles in the system. 
As an example, one can check that the degrees in $a$ and $t$ of steady state weights of all configurations computed in \eqref{wt example 2} are $4$ and $6$ respectively.

A polynomial $p(x) \in \mathbb{R}$ with $\deg_x(p) = d$ given by
\begin{equation*}
    p(x) = p_0 + p_1 x + \cdots + p_d x^d, \qquad \qquad p_d \neq 0,
\end{equation*}
is said to be \emph{palindromic} (resp. \emph{antipalindromic}) if 
$p_i = p_{d-i}$ (resp. $p_i = -p_{d-i}$) for $0 \leq i \leq d$.
Similarly, a multivariate polynomial $p(x_1, \dots, x_m)$ with degrees $d_1, \dots, d_m$ in the variables $x_1, \dots, x_m$ respectively, is said to be \emph{palindromic} (resp. \emph{antipalindromic}) if the coefficient of $x_1^{i_1} \dots x_m^{i_m}$ in $p$ is the 
same as (resp. negative of) the coefficient of $x_1^{d_1 - i_1} \dots x_m^{d_m - i_m}$
for $0 \leq i_1 \leq d_1, \dots, 0 \leq i_m \leq d_m$.
From the definition, the product of two palindromic polynomials is palindromic, the product of two antipalindromic polynomials is also palindromic, and the product of a palindromic and antipalindromic polynomial is antipalindromic.
{Similar to the definition of center of mass of a univariate palindromic polyonimal in \cite[Section 2]{ayyer-misra-2024}, 
we can define the \emph{center of mass} of any bivariate polynomial $p(x,y)$ in the variables $x$ and $y$ as the tuple 
\begin{equation}
\left( \frac{\ord_x(p)+\deg_x(p)}{2},\frac{\ord_y(p)+\deg_y(p)}{2} \right).
\end{equation}
}

We will now show that $\wta(\eta)$ is a palindromic or antipalindromic polynomial in the variables $t$ and $\pa$ according to whether $n$ is even or odd.
We have already shown in \cite[Equation (4.3)]{ayyer-misra-2024} that the $t$-binomial coefficient is palindromic as a function of $t$. Since the $t$-multinomial coefficient can be expressed as a product of $t$-binomial coefficients, it is also palindromic.
Thus, it remains to look at the product of the Pochhammer symbols.

Consider a single Pochhammer factor $(\pa; t)_m$ and look at its expansion,
\begin{equation*}
    (\pa;t)_{m} = \prod^{m-1}_{k=0} (1-\pa t^{k}) = (1-\pa)(1-\pa t) \ldots (1-\pa t^{m-1}).
\end{equation*}
There are $m$ factors, each containing two terms. Thus, each term $T$ in the expansion of this product can be represented as a binary vector $b = (b_1, \dots, b_m) \in \{0, 1\}^m$, where
$b_i = 0$ means the term $1$ is chosen from the $i$'th factor and $b_i = 1$ means the term $-\pa t^{i-1}$ is chosen. Clearly, the sign of $T$ is the parity of the number of positions $i$ such that $b_i = 1$.
Let $\bar{b} = (1 - b_1, \dots, 1 - b_m)$, with corresponding term in the expansion denoted $\overline{T}$. Then, the sign of $\overline{T}$ is the same as that of $T$ if $m$ is even and the opposite of that when $m$ is odd. Moreover, it is easy to check that 
$\deg_\pa(\overline{T}) + \deg_\pa(T) = m$ and
$\deg_t(\overline{T}) + \deg_t(T) = m(m-1)/2$.
Therefore, $(\pa;t)_{m}$ is palindromic if $m$ is even and and antipalindromic if $m$ is odd.

Now consider $\prod_{i=1}^{L}(\pa;t)_{\eta_i}$. If $n$ is even, then the number of $i$'s where $\eta_i$ is odd is also even and therefore the product becomes palindromic. Similarly, if $n$ is odd, the number of $i$'s where $\eta_i$ is odd is also odd and therefore the product becomes antipalindromic. 
Therefore $\wta(\eta)$ is palindromic if $n$ is even and antipalindromic if $n$ is odd. 
{From the calculations above, the center of mass of $\wta(\eta)$ in the variables $a$ and $t$ is
\begin{equation}
\left( \frac{n}{2},\frac{n(n-1)}{4} \right),
\end{equation}
which is the same for any configuration $\eta$. 
Therefore, the partition function is also palindromic if $n$ is even and antipalindromic if $n$ is odd. 
Since the stationary probability of $\eta$ is the ratio $\wta(\eta)/\Za_{L,n}$, is is always invariant under the transformation $a \to 1/a$ and $t \to 1/t$. }

We can check this for the example in \eqref{wt example 2}. A nice way of visualizing coefficients of two-variable polynomials is by arranging the coefficients as a matrix. For each configuration $\eta$, we write the matrix whose $(i,j)$'th entry is the coefficient of $\pa^{i-1} t^{j-1}$ below.
\[
(0,0,4) : \begin{pmatrix}
1 & 0 & 0 & 0 & 0 & 0 & 0 \\
-1 & -1 & -1 & -1 & 0 & 0 & 0 \\
0 & 1 & 1 & 2 & 1 & 1 & 0 \\
0 & 0 & 0 & -1 & -1 & -1 & -1 \\
0 & 0 & 0 & 0 & 0 & 0 & 1 
\end{pmatrix},
\]
\[
(0,1,3) : \begin{pmatrix}
1 & 1 & 1 & 1 & 0 & 0 & 0 \\
-2 & -3 & -4 & -4 & -2 & -1 & 0 \\
1 & 3 & 5 & 6 & 5 & 3 & 1 \\
0 & -1 & -2 & -4 & -4 & -3 & -2 \\
0 & 0 & 0 & 1 & 1 & 1 & 1 
\end{pmatrix},
\]
\[
(0,2,2) : \begin{pmatrix}
1 & 1 & 2 & 1 & 1 & 0 & 0 \\
-2 & -4 & -6 & -6 & -4 & -2 & 0 \\
1 & 5 & 7 & 10 & 7 & 5 & 1 \\
0 & -2 & -4 & -6 & -6 & -4 & -2 \\
0 & 0 & 1 & 1 & 2 & 1 & 1 
\end{pmatrix}
\]
and
\[
(1,1,2) : \begin{pmatrix}
1 & 2 & 3 & 3 & 2 & 1 & 0 \\
-3 & -7 & -11 & -12 & -9 & -5 & -1 \\
3 & 9 & 15 & 18 & 15 & 9 & 3 \\
-1 & -5 & -9 & -12 & -11 & -7 & -3 \\
0 & 1 & 2 & 3 & 3 & 2 & 1 
\end{pmatrix}.
\]
In each case, the matrix is invariant under a rotation of $180$ degrees. 

For comparison, we also demonstrate the antipalindromic case with odd number of particles. Consider the system with $L=3$ and $n=3$. The matrices from the steady state weights are given by
\[
(0,0,3) : \begin{pmatrix}
1 & 0 & 0 & 0 \\
-1 & -1 & -1 & 0 \\
0 & 1 & 1 & 1 \\
0 & 0 & 0 & -1 
\end{pmatrix} ,
\quad
(0,1,2) : \begin{pmatrix}
1 & 1 & 1 & 0 \\
-2 & -3 & -3 & -1 \\
1 & 3 & 3 & 2 \\
0 & -1 & -1 & -1 
\end{pmatrix},
\]
and
\[
(1,1,1) : \begin{pmatrix}
1 & 2 & 2 & 1 \\
-3 & -6 & -6 & -3 \\
3 & 6 & 6 & 3 \\
-1 & -2 & -2 & -1 
\end{pmatrix}.
\]
For all of these matrices, a rotation by $180$ degrees negates the matrix.

\section{Enriched process for $t=1$ and $\rf \in \mathbb{N}$}
\label{sec:enriched}

In this section, we define a new class of particle systems which projects to the \asip{} and when $t = 1$ and $\rf$ takes integer values. 

We will define the \emph{enriched ASIP} on configurations consisting of three kinds of objects. We have {particles depicted with integers} integers labelled $1$ through $n$, $L (\rf -1)$ identical dots $\dt$, and $L-1$ separators denoted by $\s$. Thus, each configuration has length $L + n - 1$. Further, we enforce the condition that there are $\rf-1$ $\dt$'s between any two successive separators in each configuration. 
We will think of these configurations as embedded in a circle with a separator between the first and the last object. To avoid cluttering the notation, we will not depict this separator.
We will denote by $\widehat{\Omega}^{\rf}_{L,n}$ the configuration space of the enriched ASIP.
It is not difficult to see that the cardinality of this set is
\[
|\widehat{\Omega}^{\rf}_{L,n}| =\frac{(n+L \rf -1)!}{(L \rf -1)!} = n! \, \binom{n + L \rf - 1}{n}.
\]
As an example, the set of configurations for $L=3$, $n=2$ and $\rf=2$ is
\begin{equation}\label{enriched example}
    \widehat{\Omega}^{2}_{3,\; 2} =
 \left\{
\begin{aligned}
\; \dt \s \dt \s \dt \; 1 \; 2,
\; \dt \s \dt \s \dt \; 2 \; 1,
\; \dt \s \dt \s 1 \dt 2,
\; \dt \s \dt \s 1 \; 2 \; \dt,
\; \dt \s \dt \s 2 \dt 1,
\; \dt \s \dt \s 2 \; 1 \; \dt\\
\; \dt \s \dt \; 1 \s \dt \; 2,
\; \dt \s \dt \; 1 \s 2 \; \dt,
\; \dt \s \dt \; 1 \; 2 \s \dt,
\; \dt \s \dt \; 2 \s \dt \; 1,
\; \dt \s \dt \; 2 \s 1 \; \dt,
\; \dt \s \dt \; 2 \; 1 \s \dt\\
\; \dt \s 1 \; \dt \s \dt \; 2,
\; \dt \s 1 \; \dt \s 2 \; \dt,
\; \dt \s 1 \dt 2 \s \dt,
\; \dt \s 1 \; 2 \; \dt \s \dt,
\; \dt \s 2 \; \dt \s \dt \; 1,
\; \dt \s 2 \; \dt \s 1 \; \dt\\
\; \dt \s 2 \; \dt \; 1 \s \dt,
\;\dt \s 2 \; 1 \; \dt \s \dt,
\; \dt \; 1 \s \dt \s \dt \; 2,
\; \dt \; 1 \s \dt \s 2 \; \dt,
\; \dt \; 1 \s \dt \; 2 \s \dt,
\; \dt \; 1 \s 2 \; \dt \s \dt\\
\; \dt \; 1 \; 2 \s \dt \s \dt,
\; \dt \; 2 \s \dt \s \dt \; 1,
\; \dt \; 2 \s \dt \s 1 \; \dt,
\; \dt \; 2 \s \dt \; 1 \s \dt,
\; \dt \; 2 \s 1 \; \dt \s \dt,
\; \dt \; 2 \; 1 \s \dt \s \dt\\
\; 1 \; \dt \s \dt \s \dt \; 2,
\; 1 \; \dt \s \dt \s 2 \; \dt,
\; 1 \; \dt \s \dt \; 2 \s \dt,
\; 1 \; \dt \s 2 \; \dt \s \dt,
\; 1 \; \dt \; 2 \s \dt \s \dt,
\; 1 \; 2 \; \dt \s \dt \s \dt\\
\; 2 \; \dt \s \dt \s \dt \; 1,
\; 2 \; \dt \s \dt \s 1 \; \dt,
\; 2 \; \dt \s \dt \; 1 \s \dt,
\; 2 \; \dt \s 1 \; \dt \s \dt,
\; 2 \; \dt \; 1 \s \dt \s \dt,
\; 2 \; 1 \; \dt \s \dt \s \dt
\end{aligned}
\right\}.
\end{equation}

\subsection{Dynamics}
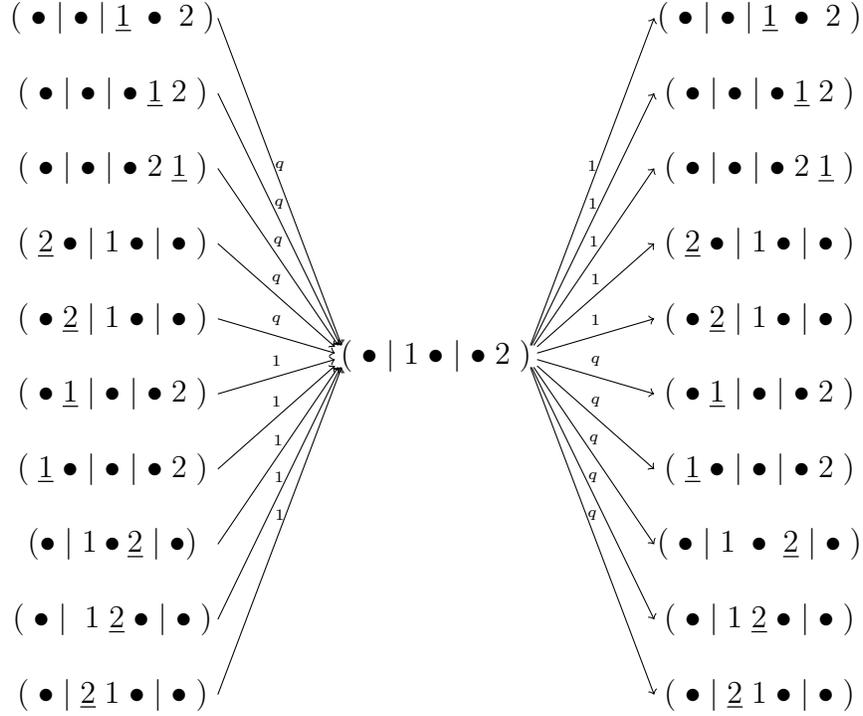
\begin{figure}[h!]\label{enriched transition example}

    \centering
    \begin{tikzpicture}
        \node(n0) at (1.2,0) {};
        \node(n99) at (-1.2,0) {};
        \node(n1) at (0,0) {$( \;\dt       \s  1 \; \dt  \s  \dt  \; 2  \;)$};
        
        \node(n2) at (4.3,4.5) {$( \; \dt   \s  \dt  \s \underline{1} \;  \dt \;  2  \; )$};
        \draw[->] (n0) -- (2.9,4.5)
        node[midway,above] {\tiny{$1$}};
        
        \node(n3) at (4.3,3.5) {$( \; \dt   \s  \dt  \s  \dt \;  \underline{1} \; 2 \; )$};
        \draw[->] (n0) -- (2.9,3.5)
        node[midway,above] {\tiny{$1$}};

        \node(n4) at (4.3,2.5) {$( \; \dt   \s  \dt  \s  \dt \; 2 \;  \underline{1} \; )$};
        \draw[->] (n0) -- (2.9,2.5)
        node[midway,above] {\tiny{$1$}};

        \node(n4) at (4.3,1.5) {$( \; \underline{2} \; \dt   \s 1 \; \dt  \s  \dt \; )$};
        \draw[->] (n0) -- (2.9,1.5)
        node[midway,above] {\tiny{$1$}};

        \node(n5) at (4.3,0.5) {$( \; \dt \; \underline{2}  \s 1 \; \dt  \s  \dt \; )$};
        \draw[->] (n0) -- (2.9,0.5)
        node[midway,above] {\tiny{$1$}};

        \node(n6) at (4.3,-0.5) {$( \; \dt  \; \underline{1} \s  \dt  \s  \dt \;  2 \; )$};
        \draw[->] (n0) -- (2.9,-0.5)
        node[midway,above] {\tiny{$q$}};

        \node(n7) at (4.3,-1.5) {$(  \; \underline{1} \; \dt   \s  \dt  \s  \dt \;  2 \; )$};
        \draw[->] (n0) -- (2.9,-1.5)
        node[midway,above] {\tiny{$q$}};

        \node(n8) at (4.3,-2.5) {$( \; \dt   \s 1 \; \dt \;  \underline{2} \s  \dt \; )$};
        \draw[->] (n0) -- (2.9,-2.5)
        node[midway,above] {\tiny{$q$}};

        \node(n9) at (4.3,-3.5) {$( \; \dt \s 1 \; \underline{2} \; \dt  \s  \dt \;   )$};
        \draw[->] (n0) -- (2.9,-3.5)
        node[midway,above] {\tiny{$q$}};

        \node(n10) at (4.3,-4.5) {$( \; \dt   \s \underline{2} \; 1 \; \dt  \s  \dt \; )$};
        \draw[->] (n0) -- (2.9,-4.5)
        node[midway,above] {\tiny{$q$}};

        \node(n11) at (-4.3,4.5) {$( \; \dt   \s  \dt  \s \underline{1} \;  \dt \;   2 \;  )$};
        \draw[->] (-2.9,4.5) -- (n99)
        node[midway,above] {\tiny{$q$}};
        
        \node(n12) at (-4.3,3.5) {$( \; \dt   \s  \dt  \s  \dt  \; \underline{1} \;  2 \;  )$};
        \draw[->] (-2.9,3.5) -- (n99)
        node[midway,above] {\tiny{$q$}};

        \node(n13) at (-4.3,2.5) {$( \; \dt   \s  \dt  \s  \dt \;   2  \; \underline{1} \;  )$};
        \draw[->] (-2.9,2.5) -- (n99)
        node[midway,above] {\tiny{$q$}};

        \node(n14) at (-4.3,1.5) {$( \;  \underline{2} \;  \dt   \s 1 \;  \dt  \s  \dt \;    )$};
        \draw[->] (-2.9,1.5) -- (n99)
        node[midway,above] {\tiny{$q$}};

        \node(n15) at (-4.3,0.5) {$( \; \dt \;  \underline{2}  \s 1  \; \dt  \s  \dt \;    )$};
        \draw[->] (-2.9,0.5) -- (n99)
        node[midway,above] {\tiny{$q$}};

        \node(n16) at (-4.3,-0.5) {$( \; \dt  \;   \underline{1} \s  \dt  \s  \dt \;   2  \; )$};
        \draw[->] (-2.9,-0.5) -- (n99)
        node[midway,above] {\tiny{$1$}};

        \node(n17) at (-4.3,-1.5) {$( \; \underline{1}  \; \dt   \s  \dt  \s  \dt \;   2  \; )$};
        \draw[->] (-2.9,-1.5) -- (n99)
        node[midway,above] {\tiny{$1$}};

        \node(n18) at (-4.3,-2.5) {$(\dt   \s 1 \dt  \underline{2} \s  \dt   )$};
        \draw[->] (-2.9,-2.5) -- (n99)
        node[midway,above] {\tiny{$1$}};

        \node(n19) at (-4.3,-3.5) {$( \; \dt   \s \;  1 \;  \underline{2} \;  \dt  \s  \dt  \;   )$};
        \draw[->] (-2.9,-3.5) -- (n99)
        node[midway,above] {\tiny{$1$}};

        \node(n20) at (-4.3,-4.5) {$( \; \dt  \s \underline{2} \;  1  \; \dt  \s  \dt  \;   )$};
        \draw[->] (-2.9,-4.5) -- (n99)
        node[midway,above] {\tiny{$1$}};
    \end{tikzpicture}
    
    \caption{All allowed incoming and outgoing transitions for state $( \; \dt   \s  1 \; \dt  \s  \dt \;   2 \; )$ with their respective rates. The particle involved in the transition is underlined.}
    \label{fig:enriched transition example}
\end{figure}

In this enriched ASIP, we let the particles hop around while the dots and separators remain stationary. 
The only hops that are allowed are those where a particle hops over a single separator to its right (resp. left) {and inserts itself between any two objects (integers, dots or separators) with  rate $1$ (resp. with rate $q$). This rate depends only on the direction of the jump and not on the objects between which the jump is made.} 
Since the enriched ASIP is periodic, particles before the first separator can jump to locations after the last separator 
with rate $q$, and similarly particles after the last separator can jump to locations before the first separator with rate $1$. 
Examples of allowed transitions are
\[
( \;\dt \; \underline{2} \s \dt \s \dt \s 1 \; \dt \;) \xrightarrow[]{q} ( \;\dt \s \dt \s \dt  \s 1 \; 2 \; \dt \;)
\]
and 
\[
( \;\dt \; 2 \s \dt \s \dt \s \underline{1 }\; \dt \;) \xrightarrow[]{1} ( \;1 \; \dt \; 2 \s \dt \s \dt \s \dt \;),
\] 
where the particle moving is underlined.
We illustrate all incoming and outgoing transitions for the chosen state $( \; \dt \s 1 \; \dt \s \dt \; 2 \; )$ in \cref{fig:enriched transition example}.

\subsection{Steady state for the enriched ASIP}
\label{sec:enriched ss}

We begin by proving that this process is ergodic. We want to show that we can go from any given state to any other state in a finite number of moves. To do this, we deploy the following algorithm. 
Starting from a given state $\hat\eta \in \widehat{\Omega}^{\rf}_{L,n}$, we can move the particle $1$ across separators until it is at the leftmost position of the configuration. We can then move the particle $2$ immediately to the right of the particle $1$, and continuing this for all the numbered particles, one obtains the configuration
\begin{equation*}
\hat{c} = (1 \;  2 \; \ldots \; n \; \dt \cdots \dt \s \dt \cdots \dt \s  \dots \s \dt \cdots \dt   ).
\end{equation*}
Now, starting from $\hat{c}$, it is easy to create any other state by sequentially sending the numbered particles starting from the highest to the lowest to their required positions in a sequence of rightward jumps as per the desired state, thus proving ergodicity. 

Since the enriched ASIP is ergodic, the steady state is unique. We now show that it is uniform, i.e.
\begin{equation}\label{enriched process weights}
    \text{wt}_{\rf, 1}(\hat\eta) = 1,
\end{equation}
for any $\hat\eta \in \widehat{\Omega}^{\rf}_{L,n}$. To prove this, it will suffice to show that the sum of the rates of incoming and outgoing transitions to any state are the same. In \cref{fig:enriched transition example} for example, this sum is $5 + 5q$.

Fix a configuration $\hat\eta$. Fix an integer $i$, $1 \leq i \leq L$. We will analyse the transitions of $\hat\eta$ that lead to changes between the $(i-1)$'th separator and the $i$'th separator, where the $0$'th separator is the one between the first and last object (which we omit in our notation).
Suppose there are $p$ particles between these separators labelled $\alpha_1,\alpha_2,\ldots,\alpha_{p}$ from left to right.
Similarly let $(\gamma_1,\gamma_2,\ldots,\gamma_{m})$ be the particles between the $(i-2)$'th separator and the $(i-1)$'th separator and $(\beta_1,\beta_2,\ldots,\beta_{r})$ be the particles between the $i$'th separator and $(i+1)$'th separator, so that $\hat{\eta}$ looks like
\begin{equation}
    \hat\eta = \ldots \underset{\underset{i-2}{\downarrow}}{\s} 
    \ldots, \gamma_1, \ldots, \gamma_{m}, \ldots 
    \underset{\underset{i-1}{\downarrow}}{\s} 
    \ldots, \alpha_1, \ldots, \alpha_{p}, \ldots  
    \underset{\underset{i}{\downarrow}}{\s} 
    \ldots, \beta_1, \ldots, \beta_{r}, \ldots  
    \underset{\underset{i+1}{\downarrow}}{\s}\ldots \; .
\end{equation}
We first analyse the forward transitions between these consecutive separators. For the outgoing transitions, each particle labelled $\alpha$ can jump to the right across the $i$'th separator. Since there are $\rf-1$ dots and $r$ particles labelled $\beta$ there, we get
$r+\rf$ available spaces to transition to. 
Since the rate for each forward transition is $1$ and there are $p$ particles labelled $\alpha$, we get the total outgoing rate in the forward direction to be
$p (r+\rf)$. Similarly, the total outgoing rate in the reverse direction is
$q p (m+\rf)$, as the rate for each reverse transition is $q$.
Now each of the outgoing transitions can be reversed with the rates flipped, so that the total incoming rate into $\hat{\eta}$ involving particles labelled $\alpha$ being 
$q p (r+\rf)$ in the reverse direction and $p (m+\rf)$ in the forward direction.

Now, let us sum the total incoming and outgoing transition rates. 
For convenience, let the number of particles between the $(i-1)$'th and $i$'th separators be $\eta_i$ for $1 \leq i \leq L$.
Then the total number of forward outgoing transitions from $\hat{\eta}$ is
\begin{equation}
\label{eq: enriched right outgoing number}
\sum_{i=1}^L \eta_i (\eta_{i+1}+\rf) = \sum_{i=1}^L \left( \eta_i \eta_{i+1} + \rf \, n \right).
\end{equation}
Similarly, the total number of forward incoming transitions into $\hat{\eta}$ is
\begin{equation}
\label{eq: enriched right incoming number}
\sum_{i=1}^L \eta_i (\eta_{i-1}+\rf) = \sum_{i=1}^L \left( \eta_i \eta_{i-1} + \rf \, n \right).
\end{equation}
The number of transitions obtained in \eqref{eq: enriched right outgoing number} and \eqref{eq: enriched right incoming number} are clearly the same due to periodic boundary conditions.
By an identical argument, one can show that the total weight of reverse outgoing transitions 
equals that of reverse incoming transitions.
Thus, the uniform distribution proposed in \eqref{enriched process weights} satisfies {the master equation (also known as the global balance equation) and is thus the stationary solution}.

\subsection{Proof of projection}
\label{sec:proj}

We will now show that the enriched ASIP defined in \cref{sec:enriched} projects to the {$(q, 1, \rf)$~ASIP}. 
To that end, define $\Pi: \widehat{\Omega}_{L,n}^{\rf} \to \Omega_{L,n}$  by
\begin{equation}
\label{defproj}
\Pi(\hat{\eta}) = (m_1, \dots, m_L), 
\end{equation}
where $m_i$ counts the number of numbered particles between $(i-1)$'th and $i$'th separator.
For the example of $\widehat{\Omega}_{3,2}^2$ shown in \eqref{enriched example}, we have
\begin{multline}
    \label{eg-proj}
\Pi^{-1}(\{ (1,0,1) \})= \{ \;
 1 \; \dt \s \dt \s 2 \; \dt \;  ,\quad \dt \; 1 \s \dt \s 2 \; \dt \; ,\quad 1 \; \dt \s \dt \s \dt \; 2 \;,\quad \dt \; 1 \s \dt \s  \dt \; 2 \;,\\
  2 \;\dt \s \dt \s 1 \; \dt \; ,\quad \dt \; 2 \s \dt \s 1 \; \dt \; ,\quad 2 \; \dt \s \dt \s \dt \; 1 \;,\quad \dt \; 2 \s \dt \s  \dt \; 1  \;
\}.
\end{multline}
We also define the rate from an enriched state to a projected state as
\begin{equation}
\label{projrate}
\textrm{rate}(\hat{\eta}\rightarrow m) = \sum_{\hat{\tau} \in \Pi^{-1}(m) } \textrm{rate}(\hat{\eta}\rightarrow \hat{\tau}).
\end{equation}

To prove this projection, we need to show
the \emph{lumping property}~\cite[Lemma~2.5]{levin_peres_wilmer.2009}
\begin{equation}
\textrm{rate}(\hat{\eta}_1 \rightarrow m') = \textrm{rate}(\hat{\eta}_2 \rightarrow m') \quad \forall \, \hat{\eta}_1,\hat{\eta}_2 \in \Pi^{-1}(m), 
\end{equation}
for all $m, m' \in \Omega_{L,n}$. 
Moreover, we have to show that this rate is the same as $\textrm{rate}(m \rightarrow m')$ for the {$(q, 1, \rf)$~ASIP}.

Let $m, m' \in \Omega_{L,n}$ such that a particle crosses the $i$'th separator 
in $m$ to reach $m'$, namely a forward transition. Then $m'_i = m_i - 1$, $m'_{i+1} = m_i + 1$, and $m'_j = m_j$ for all $j \neq i, i+1$. Let $\hat{\eta} \in \widehat{\Omega}_{L,n}^{\rf}$ such that $\Pi(\hat{\eta}) = m$. Then any transition that involves a particle between the $(i-1)$'th and $i$'th separators moving in the forward direction leading to a configuration $\hat{\eta}'$ satisfies the property that 
$\Pi(\hat{\eta}') = m'$. As we have shown in \cref{sec:enriched ss}, the total sum of these rates is $\eta_i (\eta_{i+1}+\rf)$, which is also $\textrm{rate}(m \rightarrow m')$ in the \asip{}. A similar argument goes through for reverse transitions, completing the proof of projection.

It is a standard result that if a Markov process projects onto another Markov process, then the steady state of the latter can be obtained by summing over the steady state probabilities of the former. 
In the case when $\rf$ is a positive integer, we can compute the steady state weights of the {$(q, 1, \rf)$~ASIP} using the enriched ASIP. Since the steady state weights of the enriched ASIP are  equal to $1$, $\wt(m) = \bigm| \Pi^{-1}(m)\bigm|$. To find this, we first place $m_i$ particles between the $(i-1)$'th and $i$'th separator for each $i$ in
\[
\binom{n}{m_1,\ldots,m_L}
\]
ways. Now, these $m_i$ particles have to be placed along with the $\rf - 1$ dots there.
This can be done in 
\[
\prod_{i=1}^L \frac{(m_i+\rf-1)!}{(\rf-1)!}
\]
ways. Multiplying these factors gives us the steady state weight formula found previously in \eqref{weight s.s. t=1} up to the constant $1/((\rf-1)!)^L$.

Recall that the \textit{Markov chain tree theorem}~\cite{leighton-rivest-1983,anantharam-tsoucas-1989} expresses the steady state weights as polynomials in the rates. For us, when $t = 1$, these are therefore polynomials in $\rf$ alone. 
We have so far given an alternate proof for the steady state distribution at $t = 1$ and all integer values of $\rf$, which are of course infinitely many. Since we have shown that these steady state weights coincide for infinitely many values of $\rf$, it follows that the steady state for the $(q, 1, \rf)$~ASIP given by \eqref{weight s.s. t=1} is correct as a function of $\rf$. 

\bibliography{ASIP.bib}
\bibliographystyle{alpha}
\end{document}